\documentclass[review]{elsarticle}

\usepackage{lineno,hyperref}
\modulolinenumbers[5]

\journal{Journal of Computational Sciences}

\usepackage{tikz}
\usetikzlibrary{positioning}
\usepackage{subcaption}
\usepackage{bm}
\usepackage{amsmath}
\usepackage{amssymb}
\usepackage{float}
\usepackage{svg}
\graphicspath{{figures/}}

\newcommand{\diff}[2]{\frac{\text{d}#1}{\text{d}#2}}
\newcommand{\pdiff}[2]{\frac{\partial#1}{\partial#2}}
\newcommand{\cgst}{C^T_{t_0}}

\newcommand{\revadd}[1]{#1}

\title{An improved numerical method for hyperbolic Lagrangian Coherent Structures using Differential Algebra}

\author[add1]{Jack Tyler\corref{corrauthor}}
\ead{jack.tyler@soton.ac.uk}
\cortext[corrauthor]{Corresponding author}

\author[add1]{Alexander Wittig} 
\ead{a.wittig@soton.ac.uk}

\address[add1]{Astronautics Research Group, University of Southampton, Southampton, United Kingdom, SO17 1BJ}

\begin{document}

\begin{frontmatter}

\begin{abstract}
In dynamical systems, it is advantageous to identify regions of \revadd{flow which can exhibit maximal influence on nearby behaviour}. \revadd{Hyperbolic} Lagrangian Coherent Structures have been introduced to obtain \revadd{two-dimensional surfaces which maximise repulsion or attraction }in three-dimensional dynamical systems with arbitrary time-dependence. However, the numerical method to compute them requires obtaining derivatives associated with the system, often performed through the approximation of divided differences, which \revadd{can lead} to significant numerical error and numerical noise. In this paper, we introduce a novel method for the numerical calculation of hyperbolic Lagrangian Coherent Structures using Differential Algebra called DA-LCS. As a form of automatic forward differentiation, it allows direct computation of the Taylor expansion of the flow, its derivatives, and \revadd{the eigenvectors of the associated strain tensor}, with all derivatives obtained algebraically and to machine precision. It does so without \textit{a priori} information about the system, such as variational equations or explicit derivatives. We demonstrate that this \revadd{can} provide significant improvements in the accuracy of the Lagrangian Coherent Structures identified compared to finite-differencing methods in a series of test cases drawn from the literature.
We also show how DA-LCS uncovers additional dynamical behaviour in a real-world example drawn from astrodynamics.
\end{abstract}


\end{frontmatter}

\section{Introduction}

In dynamical systems, it is often useful to identify surfaces which  separate \revadd{or maximally influence} regions of qualitatively different flow.
For time-independent systems, one often determines the geometric location of the
invariant manifolds, which partition phase space and are found by studying the system's behaviour over infinite time scales \cite{Meiss1992SymplecticTransport}. 
However, in time-\revadd{a}periodic flows, \revadd{such infinite-time behaviour is not always well defined}. Instead, the behaviour of these systems is typically studied over fixed time-scales chosen to match some practical period of interest \cite{Haller2015LagrangianStructures, Lekien2010TheManifoldsb}.

To overcome this problem, several methods for identifying \revadd{analogous structures to the invariant manifolds} in temporally aperiodic systems have been suggested. For example, one may study a number of heuristic flow diagnostics \cite{Boffetta2001DetectingTechniques, Mancho2013}, \revadd{such as the Finite-Time Lyapunov exponent (FTLE) which quantifies the separation between two trajectories which start out infinitesimally close.} However, many of these methods are only effective for simple flows and are dependent on the reference frame \cite{Haller2015LagrangianStructures}. Being heuristic, they also often lack a proper theoretical foundation as to exactly what they are indicating.

Lagrangian Coherent Structures (LCS) have been proposed to solve this problem \cite{Haller2000LagrangianTurbulence}.
A particular type of LCS, the hyperbolic LCS, is locally the most repulsive or attractive surface in a given \revadd{region of flow}, and plays an analogous role to the stable and unstable manifolds. 
Several equivalent definitions of LCS have arisen in the literature (for a review, see \cite{Hadjighasem2017ADetection}). 

A global, objective approach to the practical construction of Lagrangian Coherent Structures based on their variational theory was presented in \cite{Blazevski2014HyperbolicFlows}. 
The authors provide both the theoretical underpinning and a practical algorithm to directly construct LCS as parameterised surfaces by growing material surfaces which impose locally extreme deformation on nearby sets of initial conditions. 
These surfaces are shown to be necessarily orthogonal to certain eigendirections of the Cauchy-Green strain tensor, $\cgst$, \revadd{and further satisfy a certain criterion involving the curl of the eigenvectors of $\cgst$ to ensure the surface is locally maximally repelling or attracting}. 
This approach is valid for three-dimensional flows with general time-dependence and over arbitrarily-chosen time periods of observation.

However, there are several computational complexities associated with computing LCS using this approach \cite{Haller2015LagrangianStructures}, such as the need to account for degenerate points and \revadd{orientational} discontinuities in the eigenvector field of $\cgst$. 
More importantly, the eigenvectors of $\cgst$ must be computed precisely, yet are very sensitive to numerical errors. 
These errors are particularly troublesome near regions of intense attraction or repulsion, since large errors in $\cgst$ can quickly accumulate, yet these are also the exact regions where one would expect a hyperbolic LCS. 
The approximation of the derivatives of a flow using finite differencing is often used \cite{Farazmand2014ShearlessMaps, Haller2015LagrangianStructures,  Short2015StretchingProblems, Qingyu2020LagrangianProblem}, but this method is particularly sensitive to the grid size chosen, which must be carefully selected to account for flow behaviour over different spatial scales, which is generally difficult to determine \textit{a priori} and often selected through trial-and-error. Other such methods for approximating derivatives exist, such as the use of variational equations, where one manually derives and implements a set of adjoint differential equations that are propagated along with a reference trajectory \cite{Koon2008DynamicalDesign}. While this approach yields derivatives as accurate as the propagation along the reference trajectory, it requires the derivation, implementation and integration of $n^2$ additional equations for the first derivatives of a $n$ dimensional flow, and another $n^2(n+1)/2$ equations for the second flow derivatives. An alternative Eulerian approach for approximating $\cgst$ without the need for divided differences was presented in \cite{LEUNG20113500} by the solution of a set of partial differential equations (PDEs). However, this does not extend to the computation of the derivatives of the eigenvectors of $\cgst$ and in some cases the Eulerian approach via the solution of PDEs may be more computationally expensive than the equivalent Lagrangian approach.

Separately, Differential Algebra (DA) was originally introduced to compute high-order transfer maps for particle accelerator systems \cite{Berz1987}. 
This approach constructs a Taylor series representation of an arbitrary map in a dynamical system, and has since seen widespread use in the study of non-linearities \cite{Makino1996RemainderApplications, Makino1998RigorousAccelerators, DiMauro2015NonlinearAlgebra}, the management of uncertainties \cite{Wittig2014, Wittig2015PropagationSplitting, Massari2017, Armellin2010AsteroidApophis}, and as a form of automatic differentiation \cite{Massari2018DifferentialApplications} in a wide variety of fields. 
Unlike other numerical methods such as divided differences, the derivatives found using DA are accurate to machine precision, \revadd{and since it is a form of automatic differentiation there is no need to derive or implement any additional equations beyond the system itself. However, unlike standard automatic differentiation packages, we have additional access to a Taylor expansion about the reference point, which can be manipulated directly including by partial derivative operators (see Section \ref{sec:eigenvectorexpansion}), as suggested by the name Differential Algebra \cite{Wittig2014}}.

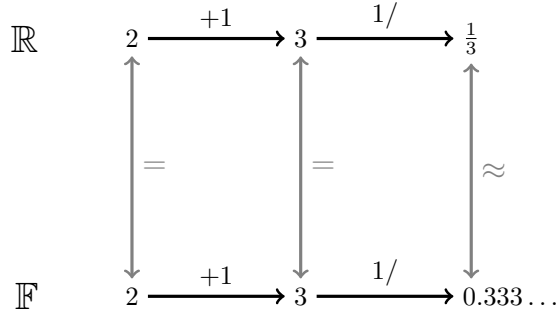
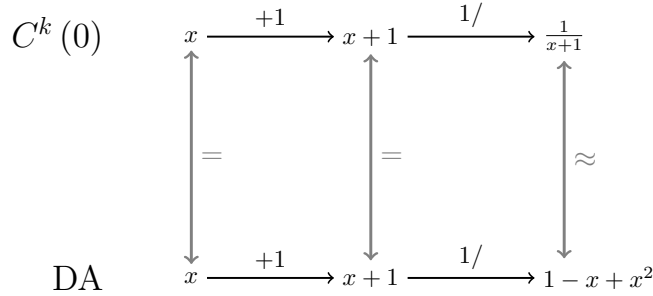
\begin{figure*}[ht]
    \centering
    \begin{subfigure}{0.75\textwidth}
        \begin{tikzpicture}
\node      (middle)                                 {};
\node      (upper)             [above=0.07\textheight of middle] {$3$};
\node      (upperright)        [right=0.2\textwidth of upper] {$\frac{1}{3}$};
\node      (upperleft)         [left=0.2\textwidth of upper]  {$2$};
\node      (upperleftleft)     [left=0.1\textwidth of upperleft]{\Large$\mathbb{R}$};
\node      (lower)             [below=0.07\textheight of middle]{$3$};
\node      (lowerleft)         [left=0.2\textwidth of lower]{$2$};
\node      (lowerleftleft)     [left=0.1\textwidth of lowerleft]{\Large$\mathbb{F}$};
\node      (lowerright)        [right=0.2\textwidth of lower]{$0.333\dots$};
\draw[->, very thick] (upperleft.east) -- (upper.west) node [midway, above] {$+1$};
\draw[->, very thick] (upper.east) -- (upperright.west) node [midway, above] {$1/$};
\draw[->, very thick] (lowerleft.east) -- (lower.west) node [midway, above] {$+1$};
\draw[->, very thick] (lower.east) -- (lowerright.west) node [midway, above] {$1/$};
\draw[<->, gray, very thick] (upperleft.south) -- (lowerleft.north) node [midway, right]{\large$\mathbf{=}$};
\draw[<->, gray, very thick] (upper.south) -- (lower.north) node [midway, right]{\large$\mathbf{=}$};
\draw[<->, gray, very thick] (upperright.south) -- (upperright.south|-lowerright.north) node [midway, right]{\large $\mathbf{\approx}$};
\end{tikzpicture}
        \caption{Evaluation of $(2+1)/3$ in the field of real numbers $\mathbb{R}$ (top) and in the floating-point approximation to $\mathbb{R}$, $\mathbb{F}$ (bottom). Each operation in $\mathbb{R}$ has a corresponding operation in $\mathbb{F}$.}
        \label{f:da_operations_floats}
    \end{subfigure}\hfill
    \begin{subfigure}{0.75\textwidth}
        \resizebox{\linewidth}{!}{
            \begin{tikzpicture}
\node      (middle)                                 {};
\node      (upper)             [above=0.07\textheight of middle] {$x+1$};
\node      (upperright)        [right=0.2\textwidth of upper] {$\frac{1}{x+1}$};
\node      (upperleft)         [left=0.2\textwidth of upper]  {$x$};
\node      (upperleftleft)     [left=0.1\textwidth of upperleft]{\Large$C^k\left(0\right)$};
\node      (lower)             [below=0.07\textheight of middle]{$x+1$};
\node      (lowerleft)         [left=0.2\textwidth of lower]{$x$};
\node      (lowerleftleft)     [left=0.1\textwidth of lowerleft]{\Large DA};
\node      (lowerright)        [right=0.2\textwidth of lower]{$1-x+x^2$};
\draw[->, thick] (upperleft.east) -- (upper.west) node [midway, above] {$+1$};
\draw[->, thick] (upper.east) -- (upperright.west) node [midway, above] {$1/$};
\draw[->, thick] (lowerleft.east) -- (lower.west) node [midway, above] {$+1$};
\draw[->, thick] (lower.east) -- (lowerright.west) node [midway, above] {$1/$};
\draw[<->, gray, very thick] (upperleft.south) -- (lowerleft.north) node [midway, right]{\large$\mathbf{=}$};
\draw[<->, gray, very thick] (upper.south) -- (lower.north) node [midway, right]{\large$\mathbf{=}$};
\draw[<->, gray, very thick] (upperright.south) -- (upperright.south|-lowerright.north) node [midway, right]{\large $\mathbf{\approx}$};
\end{tikzpicture}
        }
        \caption{Evaluation of $1/(x+1)$ in the $k-$times differentiable functions $C^k$ (top) and truncated polynomials of order $2$ represented by DA (bottom). Each operation in $C^k(0)$ has a corresponding operation in DA, approximating the resulting function in $C^k(0)$ by its Taylor expansion around $0$.}
        \label{f:da_operations_da}
    \end{subfigure}
    \caption{Comparison between the field of real numbers $\mathbb{R}$ and function space $C^k$, and their respective computer representations. The subfigures are taken from \cite{Wittig2012RigorousManifolds}.}
\end{figure*}

In this paper we introduce DA-LCS, which uses DA to improve the numerical method presented in \cite{Blazevski2014HyperbolicFlows} for determining hyperbolic LCS. 
Firstly, in Section \ref{sec:odeexpansion} we briefly review how polynomial expansions of arbitrary flows of an ordinary differential equation (ODE) can be calculated, with applications to obtaining flow derivatives of arbitrary systems to machine precision. Next, in Section \ref{sec:eigenvectorexpansion} we introduce a novel use of DA to construct algebraic expansions of the leading eigenvector of a matrix of polynomials. Both of these techniques are then combined to form the DA-LCS algorithm for computing LCS in three-dimensional flows. In Section \ref{sec:numericalexamples}, we demonstrate that this method works well in reproducing results for commonly-used `toy' problems from the literature. Lastly, in Section \ref{sec:er3bp} we present the application of DA-LCS to a more complex system from astrodynamics where the traditional method of divided differences fails to produce usable results in the literature \revadd{\cite{RosRoca2015ComputationBoundaries, Parkash2019ApplicationTrajectories}}.

\section{Mathematical background and notation}

We study the behaviour of a dynamical system
\begin{equation}\label{eq:dynamicalsystemdefinition}
    \dot{\bm{x}} = f\left(\revadd{\bm{x}},~t\right), \bm{x} \in D \subset \mathbb{R}^n,\,t \in \left[t_0,\,t_0+T\right]
\end{equation}
where $f$ is a smooth vector field considered over some time $T$ starting at time $t_0$. Denoting a trajectory of the dynamical system starting at position $\bm{x}_0$ at time $t_0$ as $\bm{x}\left(t_0,\,\bm{x}_0;\,T\right)$, the flow map of Equation \ref{eq:dynamicalsystemdefinition} is given by 
\begin{equation}
    \bm{F}^T_{t_0} : \begin{cases}
        D \rightarrow D \\
        \bm{x}_0 \mapsto \bm{x}\left(t_0,\,\bm{x}_0;\,T\right)
    \end{cases}
\end{equation}
which is assumed to be at least $k$-times continuously differentiable. The Jacobian of this flow map, $\nabla \bm{F}^T_{t_0}$, defines the right Cauchy-Green Strain Tensor (CGST) $C^T_{t_0}$, which describes the local deformation of the flow at the end of a given trajectory.
\begin{equation}\label{eq:cgstflowmap}
    \cgst = \left(\nabla \bm{F}^T_{t_0}\right)^\top \left(\nabla \bm{F}^T_{t_0}\right)
\end{equation}
with $^\top$ denoting the matrix transpose. $\cgst$ is positive-definite and symmetric, with real eigenvalues $\lambda_1 \leq \lambda_2 \leq \dots \leq \lambda_n$ and associated real eigenvectors $\bm{\zeta}_1, \bm{\zeta}_2, \dots, \bm{\zeta}_n$.

The dominant eigenvalue $\lambda_n$ can be used to calculate the finite-time Lyapunov exponent (FTLE), a measure of maximum separation of two particles advected forward under Equation \ref{eq:dynamicalsystemdefinition} that start out infinitesimally close to each other:
\begin{equation}
    \sigma^T_{t_0} = \frac{1}{2}\frac{\log{\lambda_n}}{T}.
\end{equation}
Many previous studies have leveraged the FTLE field as a heuristic indication of high regions of separation in the flow. While the FTLE has been shown to be insufficient to indicate LCS alone \citep{Haller2011}, the FTLE is a commonly-used metric and is thus used in this paper to preliminarily highlight system behaviour.

\section{Differential Algebra} 

In the following, we give a very brief introduction to Differential Algebra. For a more comprehensive treatment, the reader is referred to the literature \cite{Berz1999}.

Differential Algebra can be used as a tool to compute the derivatives of functions within a computer environment \citep{Cavenago2017On-boardNavigation, Berz1999}. Similar to how computers represent the field of real numbers as floating-point numbers, DA allows the representation and manipulation of functions in a computer \citep{wittig2017long}.

Consider two real numbers $a$ and $b \in \mathbb{R}$. The approximation to $a$ and $b$ in a computational environment is their floating-point representation $\bar{a},\,\bar{b} \in \mathbb{F}$, which essentially stores a set number of digits of its binary expansion. Any operation defined in $\mathbb{R}$, $\Box$, has a corresponding operation in $\mathbb{F},\,\boxtimes$, defined such that the result is another floating-point approximation of the operation on the real numbers $a$ and $b$, i.e. $\bar{a}\times\bar{b}$ commutes with the floating-point representation of $a\times b$, $\overline{a\times b}$.

Similarly, now consider two functions, $c$ and $d$, which are sufficiently smooth, $k-$differentiable functions of $n$ variables: $c,\,d : \mathbb{R}^n \rightarrow \mathbb{R}$. In the DA framework, a computer operates on the multivariate Taylor expansion of $c$ and $d$, $[c]$ and $[d]$, with corresponding operations to those defined in the real function space, such that the operation of $[c]\cdot[d]$ commutes with the DA representation of the product $[c\cdot d]$.

An example to demonstrate how real numbers are approximated in a computer environment is provided in Figure \ref{f:da_operations_floats} for the evaluation of the expression $1/\left(x+1\right)$ for $x = 2$ in $\mathbb{F}$ and $\mathbb{R}$. In Figure \ref{f:da_operations_floats}, we begin with $x = 2$, perform the operation $+1$ to obtain three, and then perform the operation $1/$ to compute the final expression. In $\mathbb{R}$, we obtain the solution $1/3$, and in $\mathbb{F}$ we obtain the solution $0.333\dots$ up to the limit of precision of the type. The final result of the evaluation in floating-point arithmetic is an approximation of the real computation. 

Analogously, in Figure \ref{f:da_operations_da} we evaluate the expression $1/\left(1+x\right)$ in the space $C^k\left(0\right)$ of real functions, and a DA representation of expansion order $2$. We begin with the function $c\left(x\right) = x$, and perform the operation $+1$ followed by the operation $1/$, yielding $1/\left(x+1\right)$ in the real function space, and $1 - x + x^2$ in the DA arithmetic. The result of the DA arithmetic is the Taylor expansion of $1/\left(x+1\right)$ which represents the function exactly at $x=0$, and approximates the function locally near $x = 0$ with an error of $\mathcal{O}\left(x^3\right)$. \revadd{The coefficients of the expansion are computed automatically without any further input from the user.}

Differential Algebra comprises the full set of elementary operations to efficiently operate on multivariate expansions, including operations for common intrinsic functions such as division, square roots, trigonometric functions, and exponentials, as well as operations for differentiation and integration. An important application of DA widely used in both the literature and this paper is the high-order expansion of the solution of an ODE as a function of the initial conditions \cite{Wittig2015PropagationSplitting, Armellin2010AsteroidApophis}, which is discussed in more detail in Subsection \ref{sec:odeexpansion}. In this paper, we use the Differential Algebra Computational Engine \cite{Massari2018DifferentialApplications} (DACE) to operate on polynomial expansions (`DA objects' or `DAs').

\begin{figure}
    \centering
    \includegraphics[height=.3\textheight]{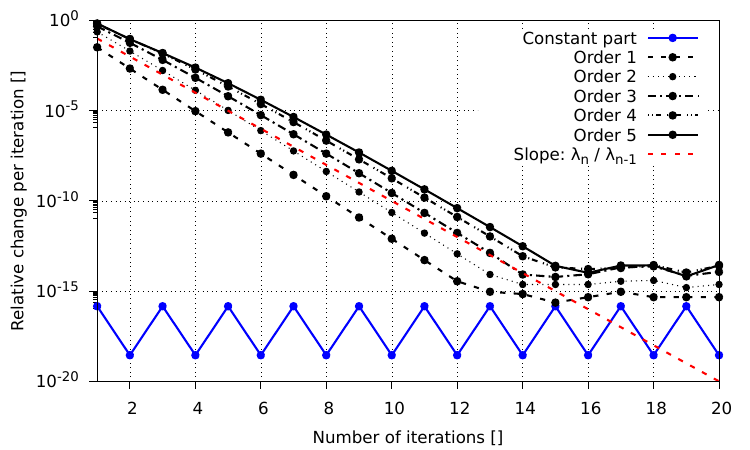}
    \caption{Relative error across all polynomial orders in successive applications of $\left[\cgst\right]$ to an initial guess containing only the floating-point dominant eigenvector at the expansion point as the constant part. Higher expansion orders (black) can all be seen converging at around the expected convergence rate $\lambda_n / \lambda_{n-1}$ (dashed red) towards the floating-point floor.}
    \label{f:eigenvector_error}
\end{figure}

\subsection{\revadd{Flow expansions to arbitrary order using Differential Algebra}}\label{sec:odeexpansion}
A key advantage of using DA is that the derivatives of flows with respect to the initial conditions can be obtained automatically and without any further effort from the user, beyond implementing the system's governing equations and the numerical integration scheme \revadd{in DA arithmetic}.

To illustrate this concept \revadd{of flow expansion}, suppose we solve the following initial value problem (IVP) numerically using a forward Euler scheme, the simplest of the Runge-Kutta family of numerical integrators
\begin{equation}
    \begin{cases}
     \bm{\dot{x}} = f\left(\bm{x},~t\right)\\
     \bm{x}\left(t_i\right) = \bm{x}_i.
    \end{cases}
\end{equation} A single step in this scheme is given explicitly by
\begin{equation}
    \bm{x}_i = \bm{x}_{i-1} + \Delta t f\left(\bm{x}_{i-1}\right)
\end{equation}
which can be expressed as a function of the initial condition $\bm{x}_0$,
\begin{align}
    \bm{x}_f =& \bm{x}_0 + \sum^n_{i = 0} hf\left(\bm{x}_0 + i\cdot h\right)
\end{align}
i.e. the initial condition is simply a sequence of operations on the initial condition, which is true for any numerical integrator of the Runge-Kutta family.

If we set $\bm{x}_0$ to be a DA representation of the initial condition by substituting the initial value with the DA identity, $\left[\bm{x}\left(t_0\right)\right] = \bm{x}\left(t_0\right) + \delta\bm{x}$, then $\bm{x}_f$ becomes a DA representation of the final condition as a function of the initial condition, $\left[\bm{x}_f
\right]$. Differentiating the polynomial thus yields the derivatives of the final condition with respect to the initial condition completely algebraically. 

As mentioned, the numerical integrator must support DA operations. Using Boost C++, which has operator overloading to operate on any type, this is relatively straightforward and its 7\textsuperscript{th}/8\textsuperscript{th} order Dormand-Prince method is used in this paper. However, care must be taken when calculating norms for error estimation in the integrator when using DA. Evaluating the usual $L_2$ norm of a vector $\lvert \bm{x} \rvert = \sqrt{\sum_{i=0}^n x_i^2}$ in DA yields another DA object representing a polynomial. As there is no ordering on the space of polynomials, this cannot be directly compared to some tolerance. Instead, we have to define the norm of a DA object which maps it into the non-negative real numbers. In this application, the norm of a DA object is taken to be the largest absolute value of any coefficient of the expansion in any order. Considering all orders in the norm allows the usual step-size control algorithms of embedded Runge-Kutta methods to control the error in all orders of the expansion, rather than just the constant part.

\subsection{Polynomial expansions of leading eigenvectors of $\cgst$ to arbitrary order}\label{sec:eigenvectorexpansion}

Since derivatives of polynomials are straight forward to compute, \revadd{we can apply the partial derivative operator to differentiating the $j-$th variable of an expansion, $\partial_j$, making it} particularly easy to assemble an expansion of $\cgst$. This means we can directly evaluate the Jacobian as
\begin{equation}
    \left[\nabla \bm{F}^T_{t_0}\right]_{ij} = \partial_j\left[\bm{x}\right]^T_{t_0,~i}
\end{equation}
from which a polynomial expansion of $\cgst$ can be assembled
\begin{equation}\label{eq:dacgst}
    \left[\cgst\right] = \left[\nabla \bm{F}^T_{t_0}\right]^\top \left[\nabla \bm{F}^T_{t_0}\right].
\end{equation}
Note that the constant part of $\left[\cgst\right]$ is the CGST at the expansion point accurate to machine precision, that is it is the same as would be approximated with via divided differences. The remaining higher order terms represent an expansion of the CGST in the neighbourhood around the expansion point.

To compute the LCS, the derivatives of the leading eigenvector of the Cauchy-Green strain tensor with respect to position are required. While divided differences can in principle again be used to obtain these derivatives, the method is susceptible to numerical noise and it is difficult to determine the most appropriate grid sizes to use. Moreover, eigenvectors are only defined up to a sign, and thus care must be taken when taking the derivatives that nearby eigenvectors have `smooth' changes in orientation.

Instead, we use a novel application of DA to obtain an expansion of the leading eigenvector of a matrix of DAs, which then can once again be differentiated directly in DA.  We simply use power (von Mises) iteration \cite{Haynsworth1966TheAnalysis.} performed in DA, which is a well-established algorithm in standard floating-point operations \cite{Chan2018IterativeEigenvalues/Eigenvectors}.

Power iteration performs the repeated evaluation of an arbitrary starting vector $\bm{b}_0$ through a matrix $A$ to obtain an approximation to its dominant unit eigenvector $\bm{b}$ through the recurrence relation
\begin{equation}\label{eq:powerlawsequence}
    \bm{b}_{m+1} = \frac{A\bm{b}_m}{\lvert\lvert A\bm{b}_m\rvert\rvert}
\end{equation}
where $\lvert\lvert \cdot \rvert\rvert$ represents a vector norm, here taken to be the $L_2$ norm, the vector $\bm{b}_0$ is an arbitrary initial vector, and $m$ is the number of iterations. The vector $\bm{b}$ will converge provided that the starting vector $\bm{b}_0$ has a nonzero component in the direction of the dominant eigenvector, and $A$ has a unique largest eigenvalue by absolute value. The theoretical convergence rate of the method between successive iterations is \revadd{the ratio of the dominant eigenvalue to the second dominant eigenvalue.}
Practically, the recurrence relation is iterated until the stopping condition $\lvert\lvert\bm{b}_{m+1}-\bm{b}_m\rvert\rvert \leq \delta$ is valid, where $\delta > 0$ is a pre-set tolerance and the norm is again taken to be an $L_2$ norm.

To convert this algorithm to DA, let $A$ now be a DA matrix with DA objects in each entry, $\left[A\right]$. Iterating it on a DA vector $[\bm{b}_0]$ will yield a DA vector $\left[\bm{b}\right]$ corresponding to the dominant eigenvector of $\left[A\right]$ with a polynomial expansion in each entry, that is it is the recurrence relation
\begin{equation}\label{eq:dapowerlaw}
    \left[\bm{b}\right]_{m+1} = \frac{\left[A\right]\left[\bm{b}_0\right]_m}{\lvert\lvert\left[A\right]\left[\bm{b}_0\right]_m\rvert\rvert}.
\end{equation}
Note that here the norm in the denominator is simply a DA evaluation of the $L_2$ (Euclidean) norm $\left[\lvert \bm{x} \rvert\right] = \sqrt{\sum_{i=0}^n \left[x\right]_i^2}$. We generalise the stopping condition from floating-point computation such that we iterate until there is no change in any order in any entry of $\left(\left[\bm{b}\right]_{m+1} - \left[\bm{b}\right]_m\right)$ above a pre-set tolerance $\delta > 0$. We set $\delta$ to be $10^{-12}$ in this paper.

To speed up convergence, and because eigenvector solvers for floating-point computations are readily available and highly efficient, we set the initial guess for $\left[\bm{b}_0\right]$ to have a constant part equal to the dominant eigenvector of the constant part of $\left[A\right]$, since we know by construction that this will be the constant part of $\left[\bm{b}\right]$.

An example of the convergence of this method is illustrated in Figure \ref{f:eigenvector_error}, which shows the maximum relative change of coefficients in $\left[\bm{b}\right]$ separated by their expansion order over repeated application of $\left[\cgst\right]$ to the initial guess of the dominant eigenvector of a trajectory in the periodic ABC flow (Section \ref{sec:periodicabc}). The theoretically expected rate of convergence $\lambda_n / \lambda_{n-1}$ can clearly be seen in the plot as a dashed red line. All orders converge at approximately the expected rate and the floating-point portion of the expression converges instantly as it was already set to the double-precision representation of the leading eigenvector.

Once the eigenvector $\left[\bm{\zeta}_n\right]$ is expanded to at least first order, the curl $\nabla\times \bm{\zeta}_n$ of the eigenvector field, which is used in the LCS construction (Section \ref{sec:lcs}), can be computed by simply applying the DA partial derivative operator \revadd{$\partial_j$} again.

To obtain the value of $\nabla\times \bm{\zeta}_n$ at the expansion point, the flow map $\bm{F}^T_{t_0}$ must be computed at least to order $2$. This is because one derivative is taken in the construction of $\cgst$ (Subsection \ref{sec:odeexpansion}), and another is then taken in $\nabla \times \bm{\zeta}_n$, both of which reduce the order of the expansion by one. 

\section{Lagrangian Coherent Structures}\label{sec:lcs}

In this section we review the method for computing LCS in three-dimensional systems given in \cite{Blazevski2014HyperbolicFlows}. For a more in-depth discussion, the reader is directed to the original paper. Once the mathematical formulation is introduced, we show how the method is computed numerically and outline the changes made from the literature in DA-LCS.

The full, three-dimensional hyperbolic LCS, which is defined as a surface that is locally maximally repelling or attracting over a given time interval $\left[t_0,\,T\right]$, is constructed from its intersections with a family of hyperplanes $\mathcal{S}$. These intersections are called \textit{reduced strainlines} and \textit{reduced stretchlines}, respectively \cite{Blazevski2014HyperbolicFlows}.

In the following, we show the mathematical formulation for repulsive LCS, whose structure is derived from the dominant eigenvector $\bm{\zeta}_3$ and whose intersections with $\mathcal{S}$ are the \textit{reduced strainlines}. A similar procedure applies to $\bm{\zeta}_1$ (reduced stretchlines) to obtain attracting LCS.

At any point $\bm{s}$ on the hyperplane, we define the reduced strainline through that point as follows: using Equation \ref{eq:dynamicalsystemdefinition}, the point is propagated from time $t_0$ to time $T$, and $\cgst$ and its eigenvectors are computed. The tangent of the reduced strainline at $\bm{s}$ is orthogonal to the leading eigenvector $\bm{\zeta}_3$ of $\cgst$ and of course also lies within the hyperplane. This is true for any point on the strainline, allowing their parameterisation to be described by the ODE
\begin{equation}\label{eq:strainlineode}
	\bm{s}^\prime = \bm{\hat{n}}_\mathcal{S} \times \bm{\zeta}_3
\end{equation}
where $\bm{\hat{n}}_\mathcal{S}$ is the unit normal to the surface at $\bm{s}$.

Strainlines which have zero \textit{helicity} $H_{\bm{\zeta}_3}$ 
\begin{equation}
	H_{\bm{\zeta}_3} = \langle\nabla\times\bm{\zeta}_3,\,\bm{\zeta}_3\rangle,
\end{equation}
where $\langle \cdot,\,\cdot \rangle$ is the inner product, are the intersections of the LCS with the reference hyperplane. The strainlines that form part of the LCS are identified by starting the integration of Equation \ref{eq:strainlineode} at initial points with zero helicity.

This analysis is repeated for each of the hyperplanes in $\mathcal{S}$. The strainlines forming part of the LCS on each hyperplane are then interpolated to produce the full 3D structure of the LCS.

To numerically implement the above procedure, we first sample points on each hyperplane in $\mathcal{S}$ on a uniformly-spaced grid and compute the helicity $H_{\bm{\zeta}_3}$ at each point. The ODE in Equation \ref{eq:strainlineode} is then rewritten in discretised form as 
\begin{equation}
	\bm{s}_i^\prime = \text{sign}\left(\bm{\zeta}_{i,\,3}\cdot\bm{\zeta}_{i-1,\,3}\right)\hat{\bm{n}}_\mathcal{S}\times\bm{\zeta}_{i,\,3}
\end{equation}
where $\bm{s}_i$ is the $i-$th point on the strainline and the term $\bm{\zeta}_{i,\,3}\cdot\bm{\zeta}_{i-1,\,3}$ is introduced to enforce continuity in the vector field by selecting the direction most closely aligned with the previous tangent vector. The selection of zero-helicity initial grid points is relaxed by allowing points where the helicity $H_{\bm{\zeta}_3}$ is below some tolerance $\alpha > 0$. The numerical integration of the ODE along the strainline continues until the sum of the helicity at each $\bm{s}_i$ divided by the number of steps performed ($i$) rises above $\alpha$. Since the eigenvector is only defined up to the sign, we integrate the strainline in both directions corresponding to $\pm\bm{\zeta}_3$ to capture the entire strainline structure.

In previous literature, divided differences was used to numerically approximate the quantities $\cgst$ and $\nabla \times \bm{\zeta}_3$ required for this procedure \cite{Farazmand2014ShearlessMaps, Haller2015LagrangianStructures,  Short2015StretchingProblems, Qingyu2020LagrangianProblem}, which can lead to significant numerical error\revadd{, particularly when computing the second derivative required for $\nabla\times\bm{\zeta}_3$}. Divided differences can either be applied on the same grid on which points are sampled, or on a finer grid used solely for the purpose of approximating the derivatives. In DA-LCS, we instead use the flow expansion described in Section \ref{sec:odeexpansion} to compute $\cgst$ \revadd{as an expansion at each grid point and around each grid point}, and the eigenvector expansion in Section \ref{sec:eigenvectorexpansion} to compute $\nabla \times \bm{\zeta}_3$ to high accuracy and without the need to alter grid sizes through trial-and-error.

The trajectories obtained through either method are segments of strainlines forming the LCS. However, since different initial points can belong to the same strainline, the trajectories often overlap. They must, therefore, be filtered to provide a single, continuous curve. Given a suitable metric $d_F$ of how close two strainline segments are, the shorter of the two strainlines is discarded whenever $d_F$ is below some threshold.

In \cite{Blazevski2014HyperbolicFlows}, this metric was the Hausdorff distance, a measure of similarity between two curves. We find that we obtain qualitatively better strainlines when using the Fr\'echet distance as a metric, which is recognised as a better measure of similarity than the Hausdorff distance in trajectory clustering problems \cite{Devogele2017, Driemel2016}. 

It is defined as follows \cite{Devogele2017}: given two curves $A$ and $B$ that are continuous mappings from $\left[0,\,1\right]$ to $\mathbb{R}^n$, define a re-parameterisation of each curve as an injective function $\Pi : \left[0,\,1\right] \mapsto \left[0,\,1\right]$, such that $\Pi\left(0\right) = 0$ and $\Pi\left(1\right) = 1$. The Fr\'echet distance $d_F$ between $A$ and $B$ is then defined with respect to their respective re-parameterisations $\Pi$ and $\Lambda$ such that
\begin{equation}
    d_F = \inf_{\Pi,~\Lambda}\,\max_{m\in\left[0,~1\right]} \lbrace d_E\left(A\left(\Pi\left(m\right)\right),~B\left(\Lambda\left(m\right)\right)\right)\rbrace
\end{equation}
where $d_E$ is the Euclidean distance. \revadd{An efficient algorithm for a numerical implementation is made available in \cite{JoaoPauloFigueira2021FastDistance}.}

\section{Arnold-Beltrami-Childress Flows}\label{sec:numericalexamples}

\revadd{To show that DA-LCS reproduces the results from the literature, w}e now apply the standard approach of divided differences and the DA-LCS method to several variations of the Arnold-Beltrami-Childress (ABC) flow, as studied in \cite{Blazevski2014HyperbolicFlows}. For each example, we present the equations of motion, the FTLE field, the helicity field, and the resulting strainlines. The results obtained using divided differences each use the manually-determined optimal grid size for each application that produces the qualitatively `best' results, to allow for a fair comparison. \revadd{Grid sizes between $0.1$ and $5$ times the nominal grid size were analysed.} No such adjustments are needed when using DA-LCS.

\subsection{Steady Arnold-Beltrami-Childress flow}\label{sec:steadyabcflow}

\begin{figure}
    \centering
    \includegraphics[height=.3\textheight]{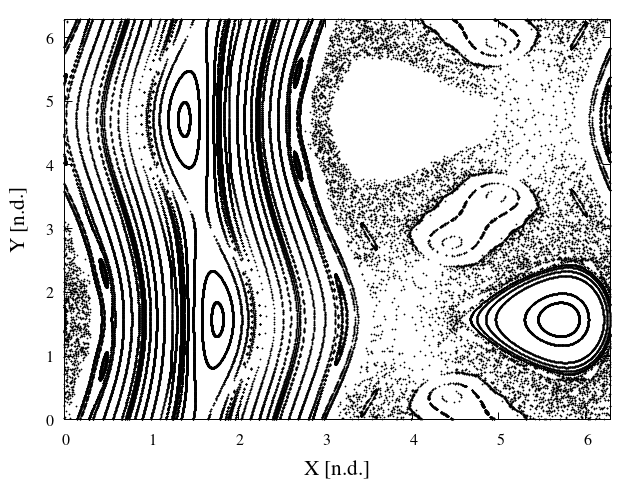}
    \caption{Poincar\'e section (return map) for the steady ABC flow on the $z = 0$ plane; generated using a $15 \times 15$ grid of initial points with integration time $T = 1500$.}
    \label{f:steady_abc_poincaire}
\end{figure}

\begin{figure}
\centering
\begin{subfigure}{0.75\textwidth}
    \centering
    \includegraphics[width=\linewidth]{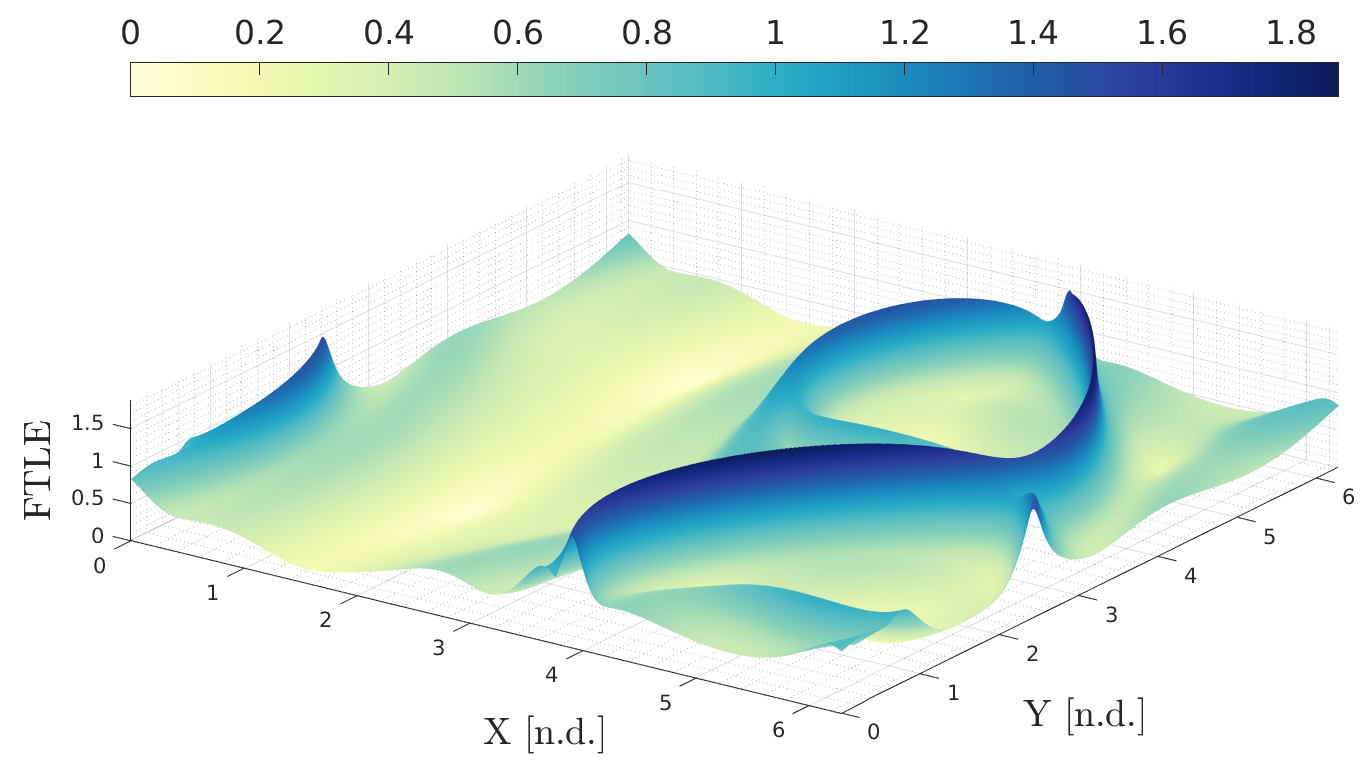}
    \caption{FTLE field obtained using DA-LCS.}
    \label{f:steady_abc_ftle_da}
\end{subfigure}

\begin{subfigure}{0.75\textwidth}
    \centering
    \includegraphics[width=\linewidth]{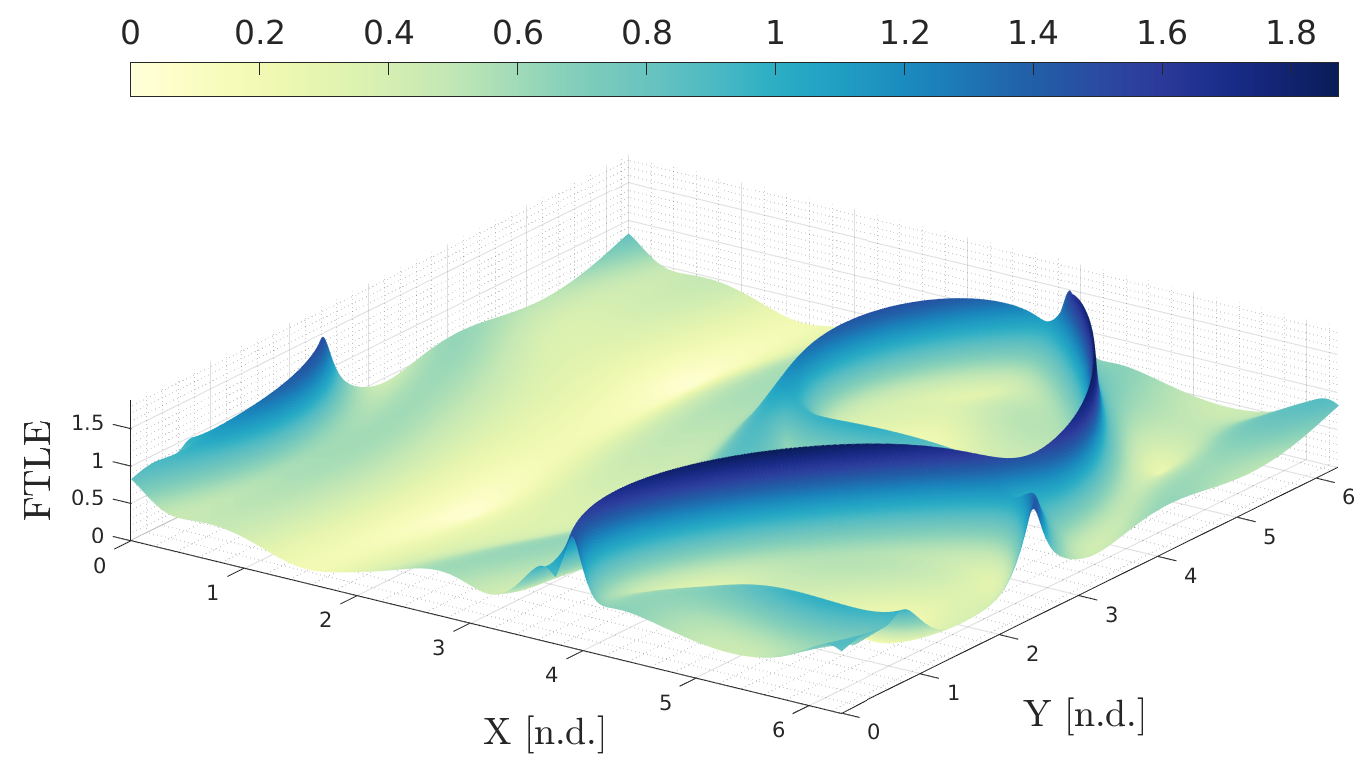}
    \caption{FTLE field obtained using divided differences with auxiliary grid spacing of $0.05$ of the nominal grid spacing in all directions.}
    \label{f:steady_abc_ftle_double}
\end{subfigure}
\caption{Finite-time Lyapunov field for the steady ABC flow from $t = 0$ to $T = 3$ using DA-LCS and divided differences. The fields strongly agree, suggesting that computing $C^3_0$ using divided differences is not a major source of error in this example.}
\label{f:steady_abc_ftle}
\end{figure}

\begin{figure*}
    \centering
    \begin{subfigure}{0.75\textwidth}
        \includegraphics[width=\textwidth]{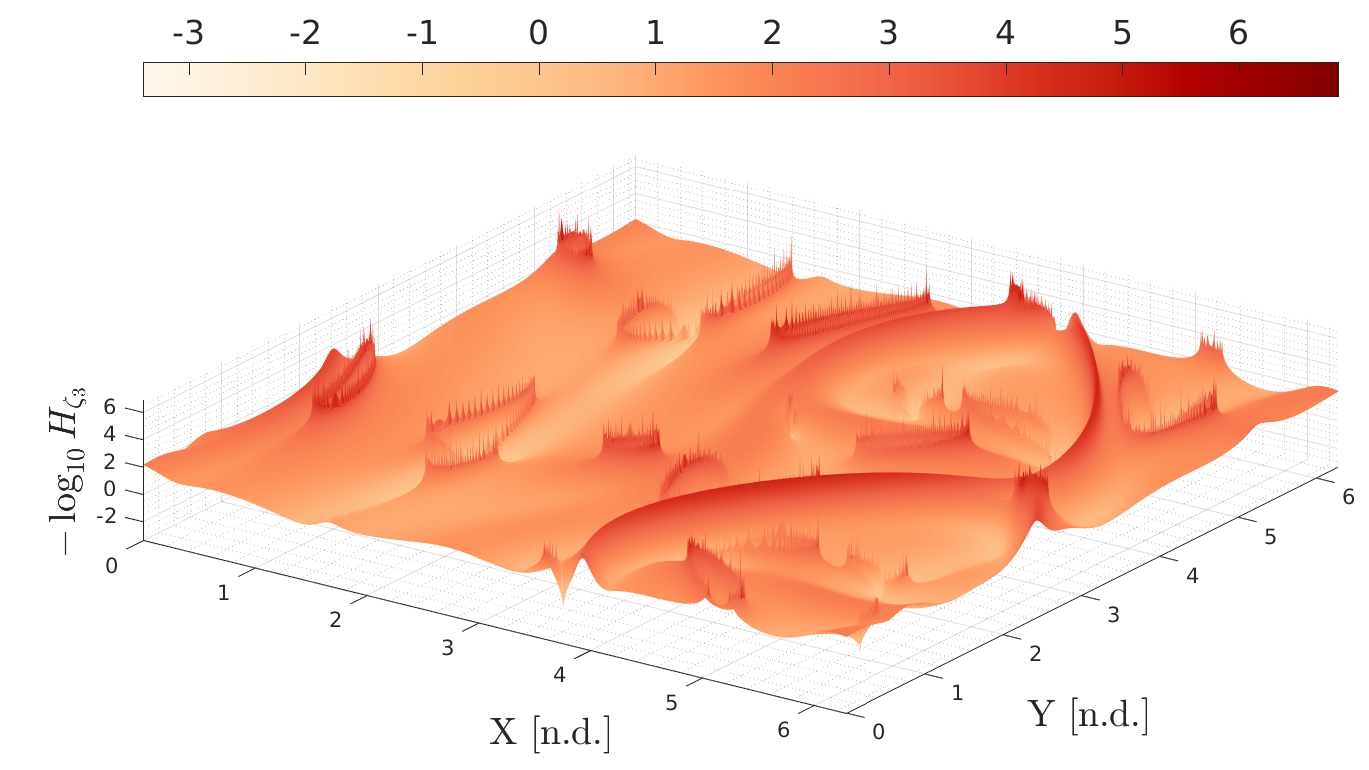}
        \caption{Helicity field obtained using DA-LCS.}
        \label{f:helicity_field_steady_da}
    \end{subfigure}
    
    \begin{subfigure}{0.75\textwidth}
        \includegraphics[width=\textwidth]{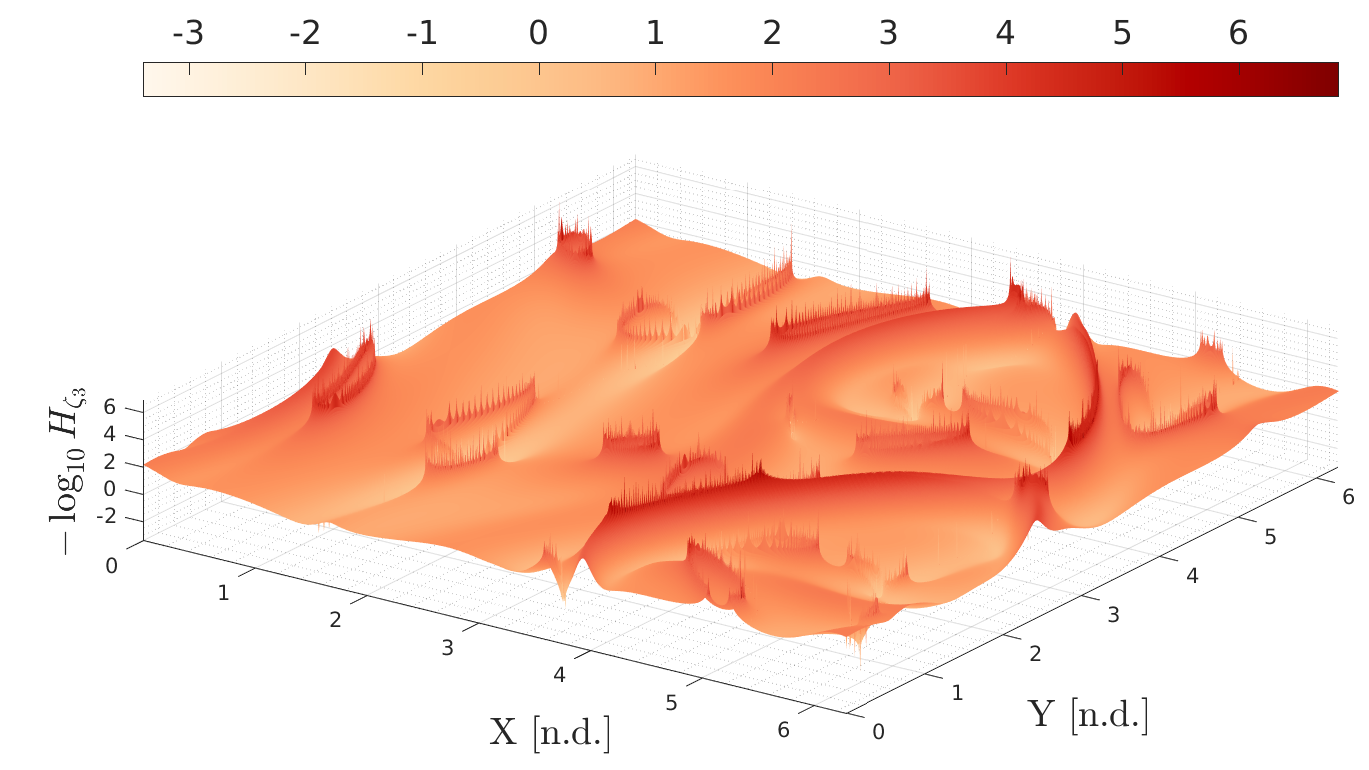}
        \caption{Helicity field obtained using divided differences with auxiliary grid spacing of $0.05$ of the nominal grid spacing in both $x$ and $y$. The same grid is used in computing both $C^3_0$ and $\nabla\times\bm{\zeta}_3$.}
        \label{f:helicity_field_steady_dd}
    \end{subfigure}
    \caption{$-\log{H_{\zeta_3}}$ for the steady ABC flow from $t_0 = 0$ to $T = 3$ using DA-LCS and divided differences. Again both strongly agree, showing that DA-LCS is working. The DA-LCS structure is a little smoother along the main ridge on the right compared to divided differences, \revadd{making the identification of seed points} more robust.}
    \label{f:steady_abc_helicity_field}
\end{figure*}

\begin{figure}
    \centering
    \includegraphics[height=.3\textheight]{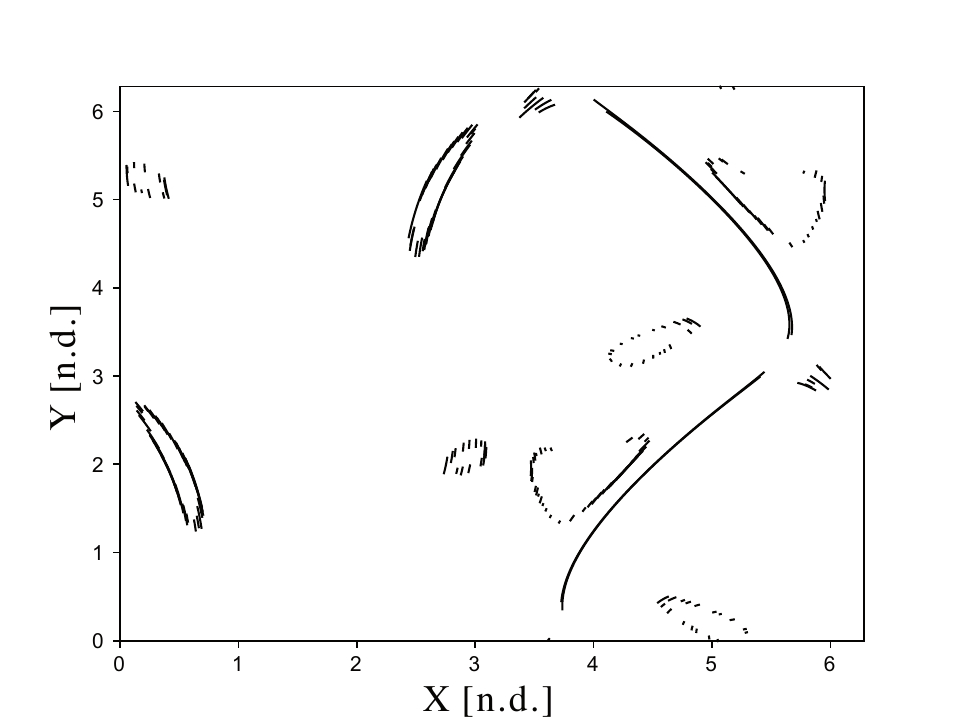}
    \caption{Final, filtered strainlines for the steady ABC flow on the $z=0$ plane computed using DA-LCS. The structure is formed of approximately $240$ strainline segments.}
    \label{f:abc_final_strainlines}
\end{figure}

We first consider the steady Arnold-Beltrami-Childress flow, using the problem parameters and reference planes presented in \cite{Blazevski2014HyperbolicFlows}. The ABC flow is an exact solution to Euler's equation, and its equations of motion in Cartesian coordinates are
\begin{eqnarray}\label{eq:steadyabceomstart}
    \dot{x} &=& A\sin{z} + C\cos{y} \\
    \dot{y} &=& B\sin{x} + A\cos{z} \\
    \dot{z} &=& C\sin{y} + B\cos{x}\label{eq:steadyabceomend}
\end{eqnarray}
\revadd{with} parameter values $A = \sqrt{3},\,B = \sqrt{2},\,C = 1.0$. 
To illustrate the behaviour of this system, the Poincar\'e section in the $x$-$y$ plane is shown in Figure \ref{f:steady_abc_poincaire}, computed from a regular $15\times 15$ grid of initial points and an integration time of $T = 1500$.

For the LCS computation, matching previous literature the set of reference planes are taken to be
$$\mathcal{S} = \lbrace \left(x\,, y\,, z\right) \in \left[0,\,2\pi\right]^3~:~z\in \{0, 0.005, 0.01 \dots, 0.1\}\rbrace,$$
that is the $x$-$y$ plane evenly spaced along the $z$ axis.
However, within each plane we alter the grid size used. \cite{Blazevski2014HyperbolicFlows} use a $500\times500$ grid on which to compute the underlying helicity field, and then sample seed points for the ODE in Equation \ref{eq:strainlineode} on a reduced grid of $600\times 10$. \revadd{While the authors acknowledge that sampling every point on a dense grid is numerically inefficient, to simplify analysis, ensure we capture all of the flow's behaviour, and to work off of the assumption of no \textit{a priori} knowledge} we perform all stages of the analysis on a $1000\times1000$ grid defined for each hyperplane in $\mathcal{S}$. \revadd{In practice, additional information about the system may be available to search more efficiently for LCS seed points, such as searching on a fixed line or only in a certain region of flow.}

The system defined by Equations \ref{eq:steadyabceomstart}-\ref{eq:steadyabceomend} is integrated forward for $3$ non-dimensional time units using the DA-compatible numerical integrator introduced previously, with an integration tolerance of $10^{-13}$. A helicity tolerance of $\alpha = 10^{-4}$ is applied to determine seed points and terminate the numerical integration. A minimum distance of $d_F = 0.04$ is used in the strainline segment filtering. These parameters are chosen from visual examination of the helicity field and resulting strainline structure for all of the examples in this paper.

The FTLE fields on the $z = 0$ plane for this flow, computed using DA-LCS and divided differences, are shown in Figures \ref{f:steady_abc_ftle_da} and \ref{f:steady_abc_ftle_double}, respectively. The two FTLE fields are very similar, which suggests that the computation of $C^3_0$ and its dominant eigenvalue agrees across the two methods.

In the DA-LCS and divided difference helicity fields on the $z = 0$ plane, shown in Figures \ref{f:helicity_field_steady_da} and \ref{f:helicity_field_steady_dd} respectively, some first differences can be seen. While the two methods qualitatively agree on the structure of the field, the DA-LCS method produces smoother peaks and ridges in the field for the primary features in the flow. This is particularly visible on the main ridge in the bottom right corner around $X=4$ and $Y=1$. \revadd{This makes the identification of seed points in the flow more straightforward.}

\begin{table}[]
    \centering
    \begin{tabular}{p{0.2\textwidth}p{0.2\textwidth}p{0.2\textwidth}p{0.2\textwidth}}
    \hline\textbf{Method} & \textbf{Time to compute} $H_{\bm{\zeta}_3}$ \textbf{field [s]} & \textbf{Time to compute 100 strainlines [s]} & \textbf{Average function evaluations per unit length} \\\hline
    Divided differences & 224.902 & 4684.689 & 7498.171 \\
    DA-LCS & 611.360 & 1108.571 & 78.392\\\hline
    \end{tabular}
    \caption{\revadd{Core time required to compute the LCS on one reference plane for the steady ABC flow using divided differences and DA-LCS on Intel Xeon E5-2670 processors. While DA-LCS is slower to determine the initial $H_{\bm{\zeta}_3}$ field, it is quicker at the integration of a representative set of strainlines and can grow much longer strainlines with the same number of evaluations of Equation \ref{eq:strainlineode} as divided differences. Importantly, divided differences requires significant grid size tuning, which may make the time required to determine $H_{\bm{\zeta}_3}$ slower overall when used practically.}}
    \label{tbl:compute_times}
\end{table}

The resulting strainlines on the $z = 0$ plane for this flow are shown in Figure \ref{f:abc_final_strainlines}, and follow the expected structure from the helicity field presented in Figure \ref{f:helicity_field_steady_da}. 
We note the existence of several `loops' in the helicity field, particularly in the left-hand side of the field. The strainline segments at these points grow transverse to the ridges at certain points, and do not track along the ridge as would be expected. This behaviour is also present when computing LCS with divided differences. These small strainline segments are not present in \cite{Blazevski2014HyperbolicFlows} due to being missed by the largely reduced $600\times 10$ grid resolution used there. This explains their omission from the literature, and we do not investigate this issue further here, although we note the existence of similar structure in \cite{Palmerius2009FlowSurfaces}.

The total strainline structure in Figure \ref{f:abc_final_strainlines} for this test case is formed of approximately $240$ individual strainline segments. We remark that the distribution of the number of strainlines with respect to their length is largely bimodal. The `loops' discussed previously contain lots of short segments, while the main wishbone-like structures are formed from only several long strainlines. This distribution of the number of strainline segments with respect to their length is similar across all test cases studied here that are variations of the ABC flow.

\revadd{We now discuss the computational and numerical performance of DA-LCS, using the steady ABC flow as an example. The total time to compute the full LCS on 48 2.0GHz Intel Xeon E5-2670 processors is given in Table \ref{tbl:compute_times}, broken down by the time required to obtain the initial $H_{\bm{\zeta}_3}$ field and then a representative set of $100$ strainlines. The set of 100 strainlines is chosen to be the 100 points with lowest $H_{\bm{\zeta}_3}$, integrated until the running average of helicity rises above $10$ times the initial value. Visual inspection of the initial conditions confirms that the seed points are sufficiently `close' in both divided differences and DA-LCS that they are assumed to represent the same behaviour.}

\revadd{We find that DA-LCS is slower than divided differences for computing the initial helicity field since two orders are computed, requiring more CPU instructions per operation, and because fewer optimisations can be made by the compiler compared to native double-precision types. However, since DA-LCS requires no tuning of grid size, this computational deficit is eliminated as soon as more than two trial computations of the LCS using divided differences has to be performed to obtain the `optimal' grid size in every dimension. Moreover, owing to better numerical performance, the strainline integration is approximately four times faster using DA-LCS than using divided differences, since the integrator can take larger steps than with divided differences while still controlling the error in the integration of Equation \ref{eq:strainlineode}. We also find that the strainlines obtained with DA-LCS are on average $10$ times longer than when using divided differences for the representative set here; this may mean that more sophisticated search methods for identifying seed points, such as the method of searching on a fixed line mentioned earlier, would be more feasible in DA-LCS. Both improvements in strainline integration are due to the elimination of numerical noise introduced by divided differences, which is not present in DA.}

\begin{figure*}[htbp]
\centering
    \begin{subfigure}{0.75\textwidth}
    \centering
    \includegraphics[width=\linewidth]{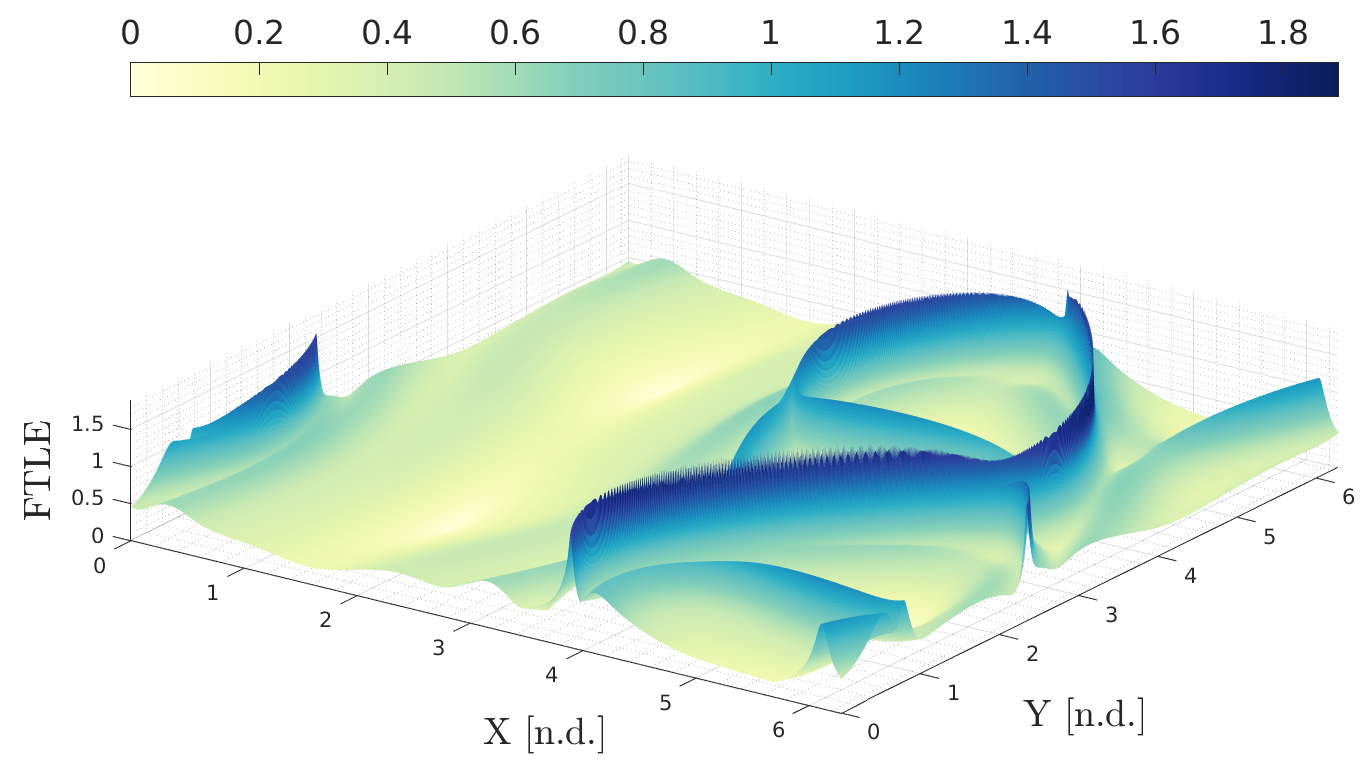}
    \caption{Computed using DA-LCS.}
    \label{f:periodic_abc_ftle_da}
    \end{subfigure}
    
    \begin{subfigure}{0.75\textwidth}
    \centering
    \includegraphics[width=\linewidth]{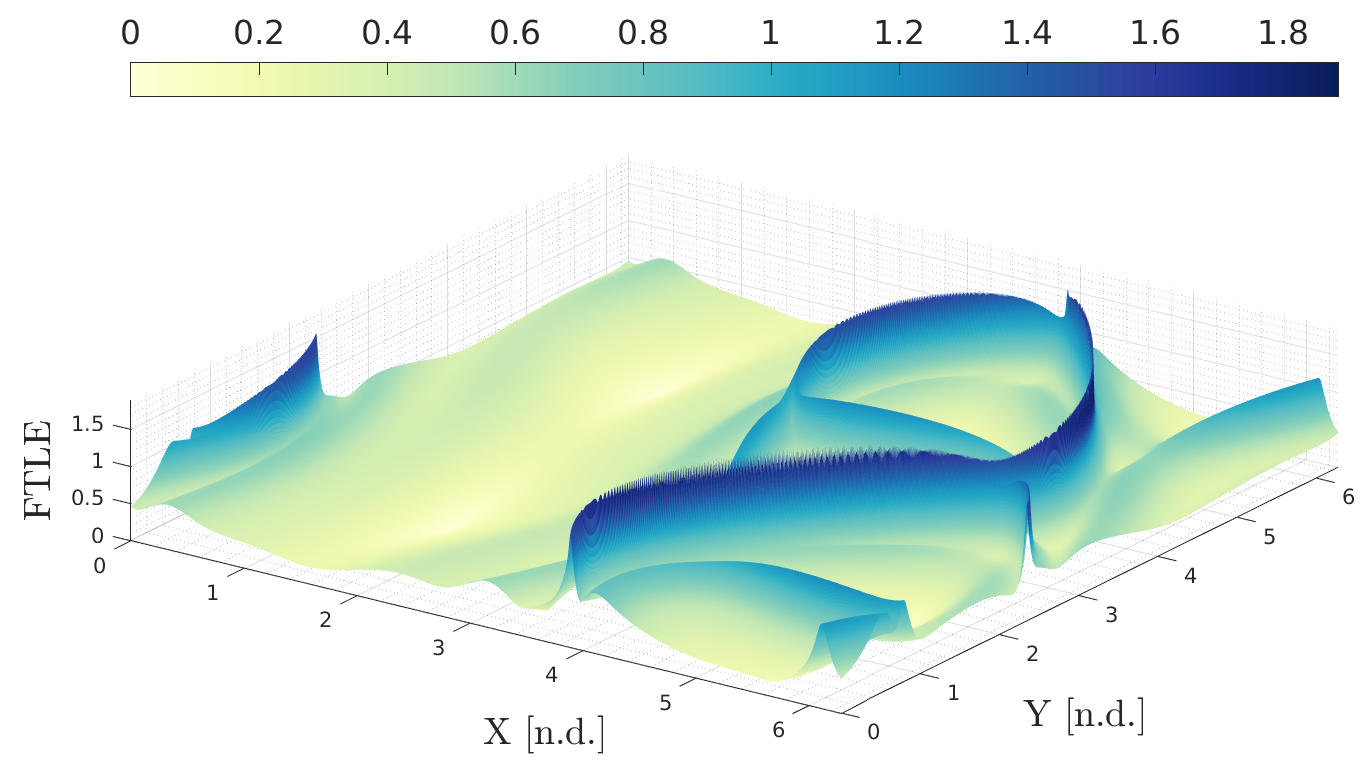}
    \caption{FTLE field obtained using divided differences with auxiliary grid spacing of $0.05$ of the nominal grid spacing in all directions.}
    \label{f:periodic_abc_ftle_dd}
    \end{subfigure}
    \caption{Finite-time Lyapunov exponent field for the periodic ABC flow from $t_0 = 0$ to $T = 4.0$, obtained using DA-LCS and divided differences. Again, the FTLE field agrees between divided differences and DA-LCS, suggesting divided differences on the correct auxiliary grid in this case accurately approximates $C^4_0$.}
\end{figure*}

\begin{figure*}[htbp]
    \centering
    \begin{subfigure}{0.75\textwidth}
        \includegraphics[width=\linewidth]{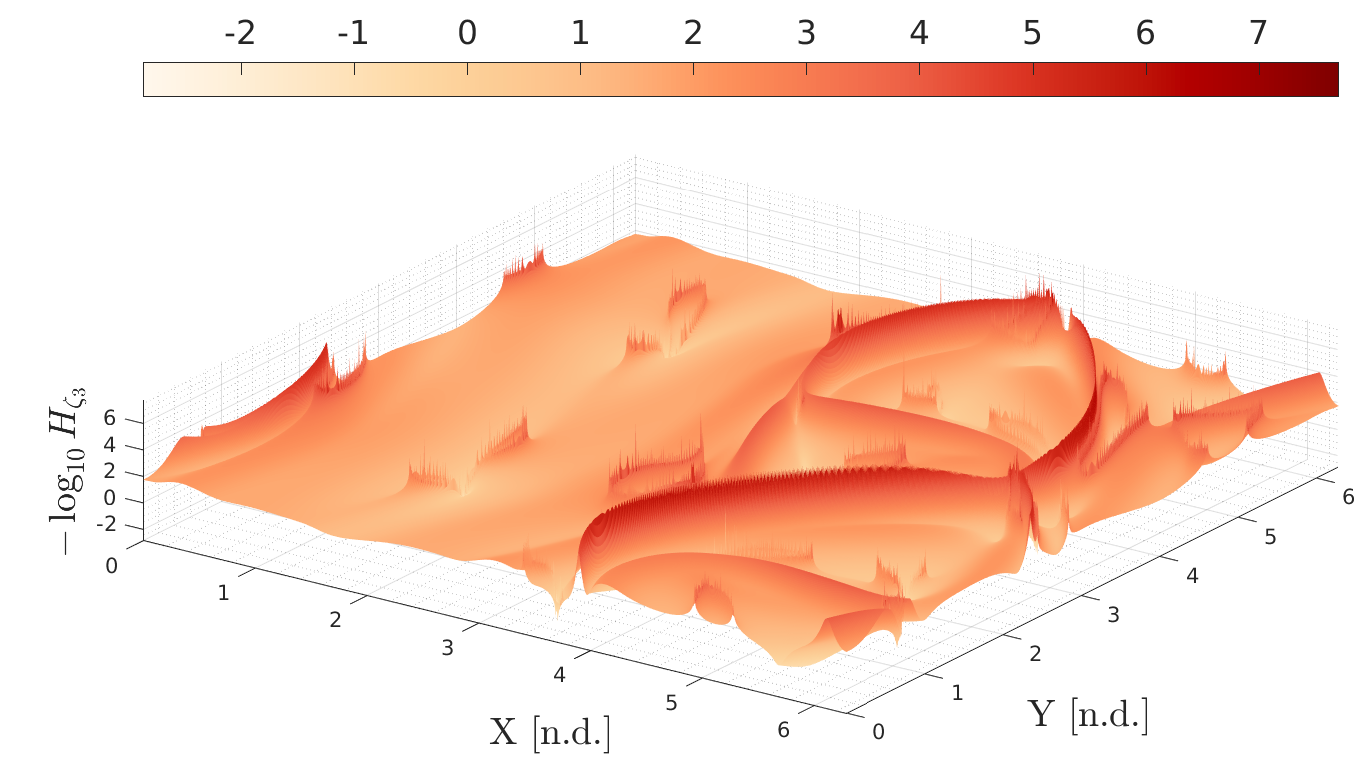}
        \caption{Computed using DA-LCS.}
        \label{f:periodic_abc_helicity_da}
    \end{subfigure}
    
    \begin{subfigure}{0.75\textwidth}
        \includegraphics[width=\linewidth]{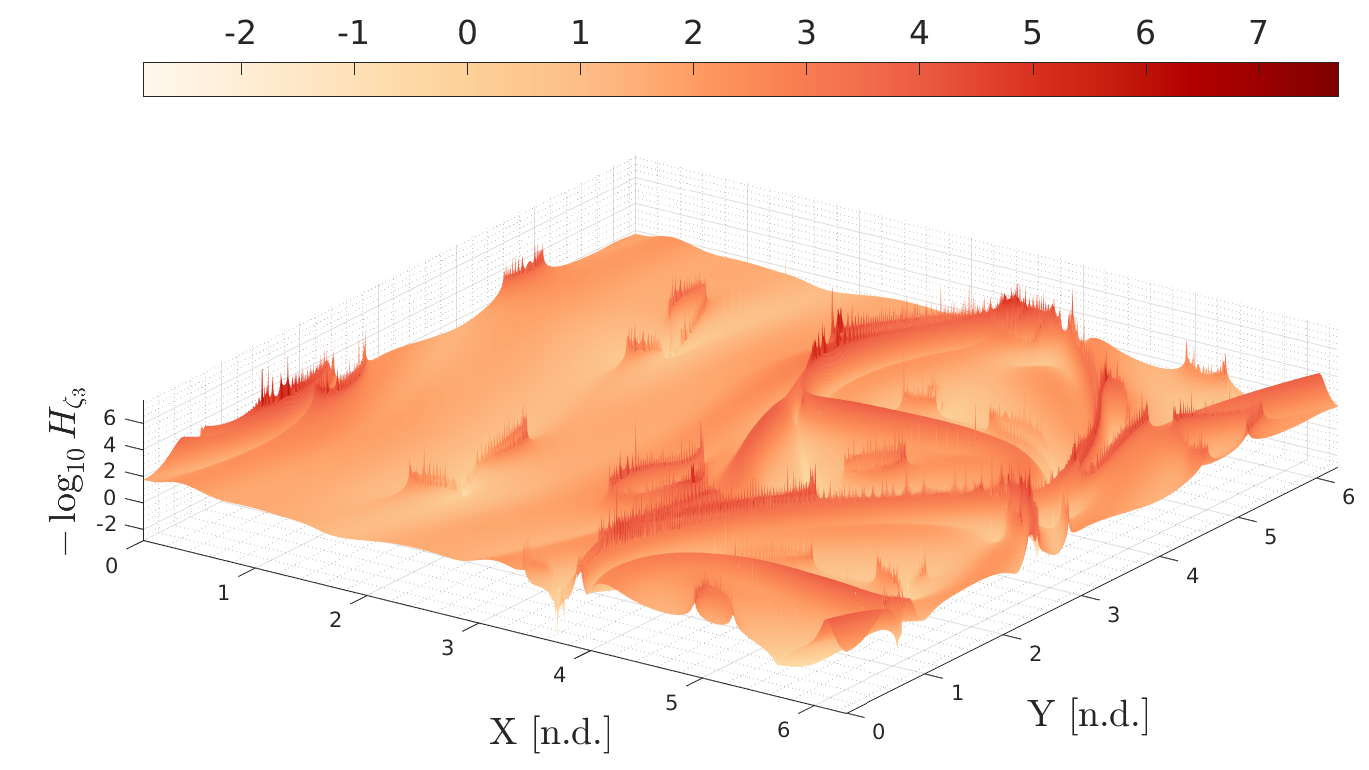}
        \caption{Helicity field obtained using divided differences with auxiliary grid spacing of $0.05$ of the nominal grid spacing in all directions. The same grid is used in computing both $C^4_0$ and $\nabla\times\bm{\zeta}_3$.}
        \label{f:periodic_abc_helicity_dd}
\end{subfigure}%
\caption{$-\log H_{\bm{\zeta}_3}$ for the periodic ABC flow computed using DA-LCS and divided differences from $t_0 = 0$ to $T = 4.0$. Here DA-LCS highlights in particular the main ridge on the right more clearly and smoothly than divided differences.}
\end{figure*}

\subsection{Periodic Arnold-Beltrami-Childress Flow}\label{sec:periodicabc}

We now consider a time-periodic version of the Arnold-Beltrami-Childress flow with equations of motion
\begin{eqnarray}\label{eq:periodicabceomstart}
    \dot{x} &=& \left(A + 0.1\sin{t}\right)\sin{z} + C\cos{y} \\
    \dot{y} &=& B\sin{x} + \left(A + 0.1\sin{t}\right)\cos{z} \\
    \dot{z} &=& C\sin{y} + B\cos{x}\label{eq:periodicabceomend}.
\end{eqnarray}

The hyperplanes $\mathcal{S}$ and grids are the same as in the case of the steady ABC flow, but now with integration times $t_0 = 0$ and $T = 4$ to again match the literature exactly. A helicity tolerance of $\alpha = 5\times10^{-5}$ is used, with a distance threshold $d_F = 0.02$.

Mirroring the analysis in the steady case, the FTLE fields for both DA-LCS and divided differences are shown in Figures \ref{f:periodic_abc_ftle_da} and \ref{f:periodic_abc_ftle_dd}, respectively. Again, there is little qualitative difference between the two fields. The differences in smoothness in the helicity fields are, however, more pronounced between Figures \ref{f:periodic_abc_helicity_da} and \ref{f:periodic_abc_helicity_dd}. The main wishbone-like structure is particularly `spiky' when using divided differences. With DA-LCS, there is a smooth, well-defined ridge of consistently low helicity for the algorithm to detect with much lower numerical noise; in fact, our helicity threshold is approximately two orders of magnitude lower than used in literature but recovers qualitatively similar structures.

\begin{figure}
    \centering
    \includegraphics[height=.3\textheight]{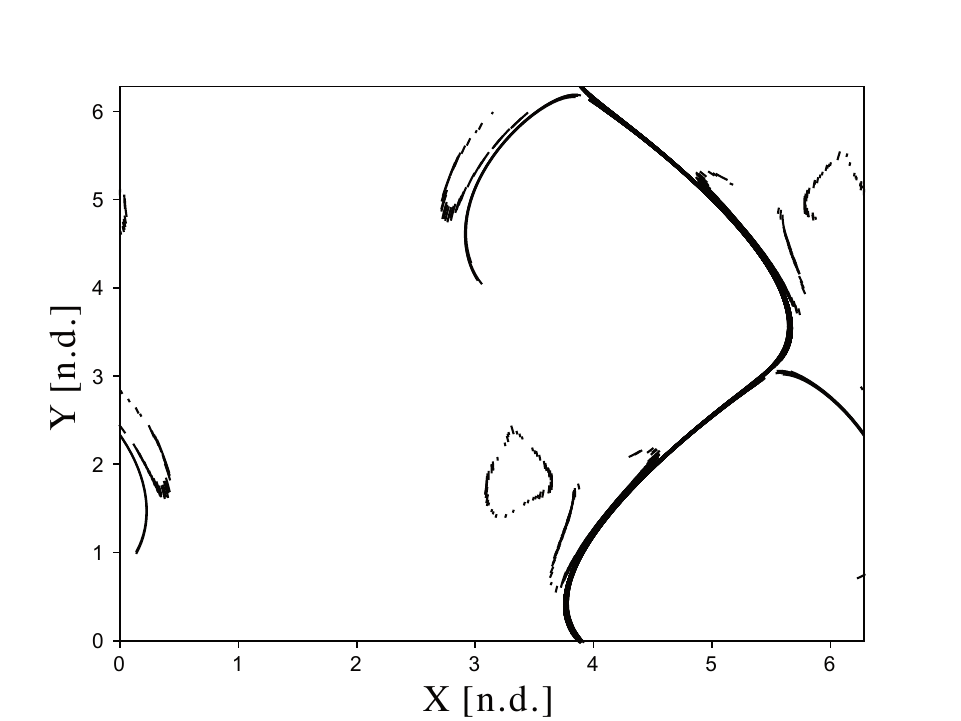}
    \caption{Final strainlines for the periodic ABC flow on the $z = 0$ plane computed using DA-LCS, after filtering. The strainline structure is composed of approximately $250$ strainline segments.}
    \label{f:periodic_abc_final_strainlines}
\end{figure}

Finally, the strainlines on the $z = 0$ plane for this system computed using DA-LCS are shown in Figure \ref{f:periodic_abc_final_strainlines}. Approximately $250$ strainline segments determine the full strainline structure on the $z = 0$ plane for this example. As with the steady ABC flow, the distribution of the number of strainlines with respect to their length is largely bimodal, and the majority of these segments are found in the `loops' in the strainline structure.

\subsection{Chaotically-forced Arnold-Beltrami-Childress flow}\label{sec:chaoticabc}
  
\begin{figure*}
\centering
\begin{subfigure}{0.75\textwidth}
\includegraphics[width=\linewidth]{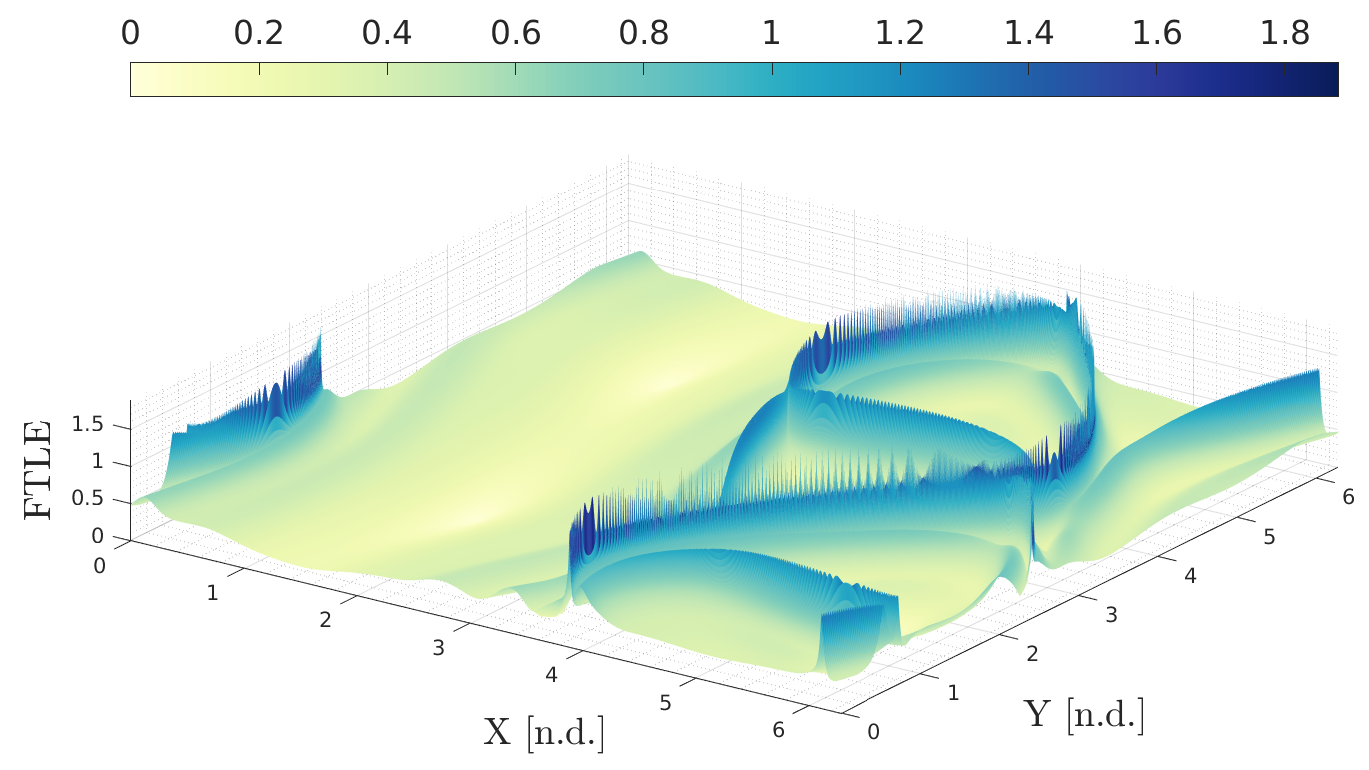}
\caption{Computed using DA-LCS.}
\label{f:forced_abc_ftle_da}
\end{subfigure}
\begin{subfigure}{0.75\textwidth}
\includegraphics[width=\linewidth]{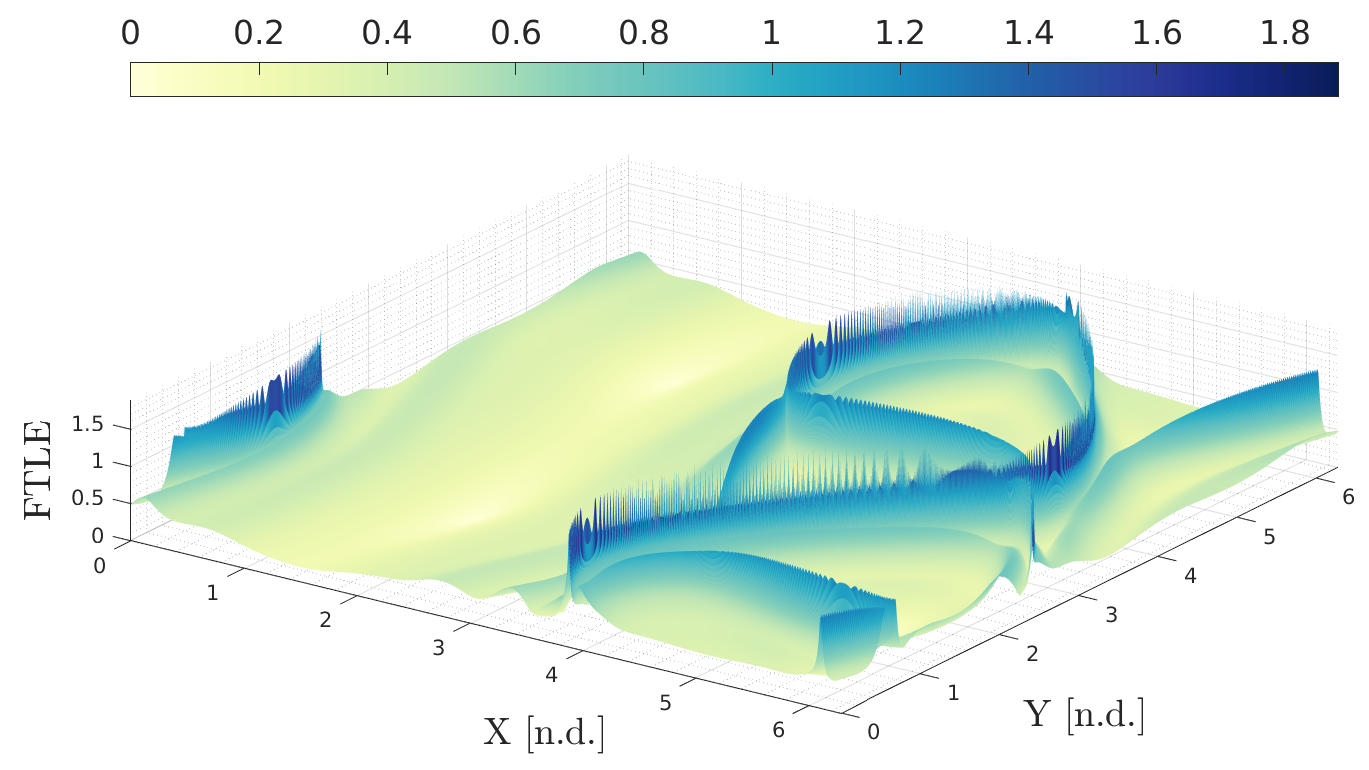}
\caption{FTLE field obtained using divided differences with auxiliary grid spacing of $0.1$ of the nominal grid spacing in all directions.}
\label{f:forced_abc_ftle_dd}
\end{subfigure}
\caption{Finite-time Lyapunov exponent field for the chaotically-forced ABC flow and an integration time from $t_0 = 0$ to $T = 5.0$. The fields still strongly agree, suggesting that again the computation of $C^5_0$ on the optimal auxiliary grid is not a major source of error for this example.}
\label{f:chaotic_abc_ftle}
\end{figure*}

\begin{figure*}
\centering
\begin{subfigure}{0.75\textwidth}
\includegraphics[width=\linewidth]{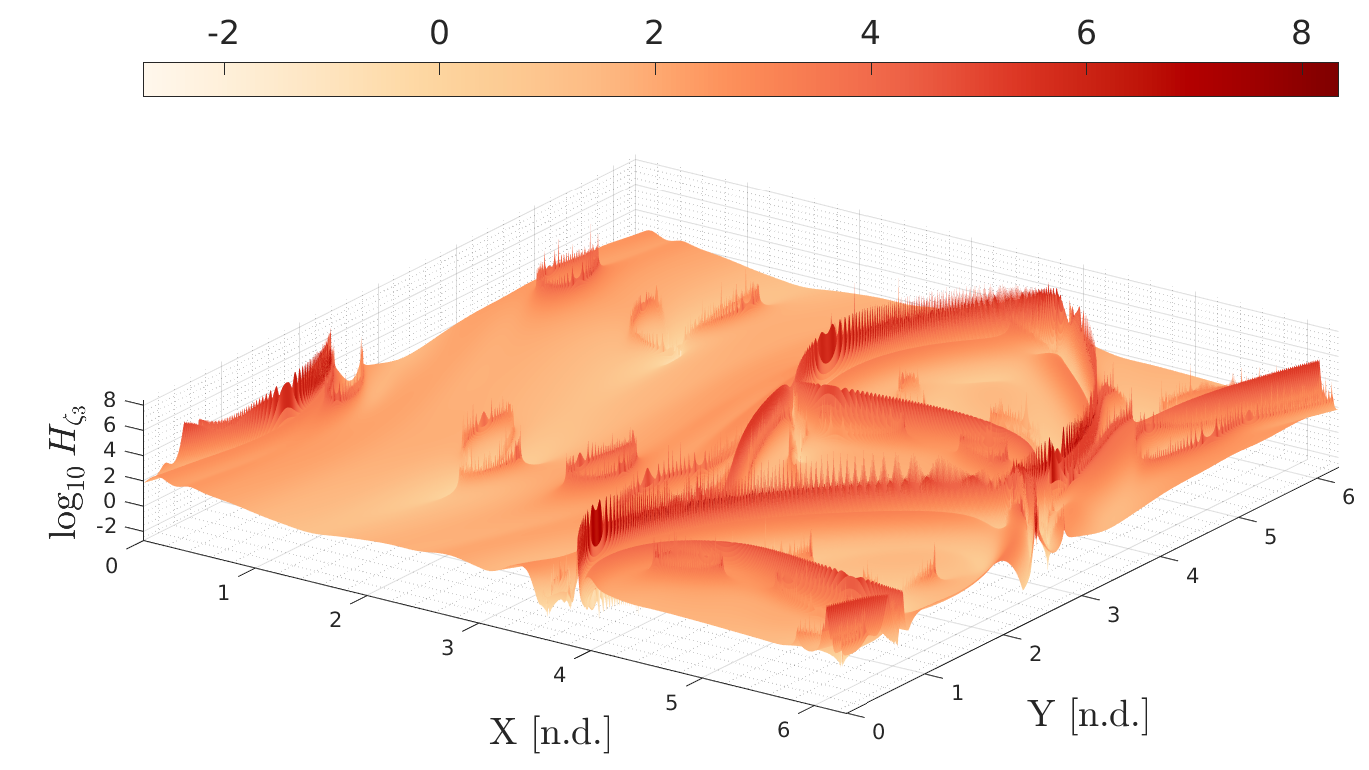}
\caption{Computed using DA-LCS.}
\label{f:forced_abc_helicity_da}
\end{subfigure}
\begin{subfigure}{0.75\textwidth}
\includegraphics[width=\linewidth]{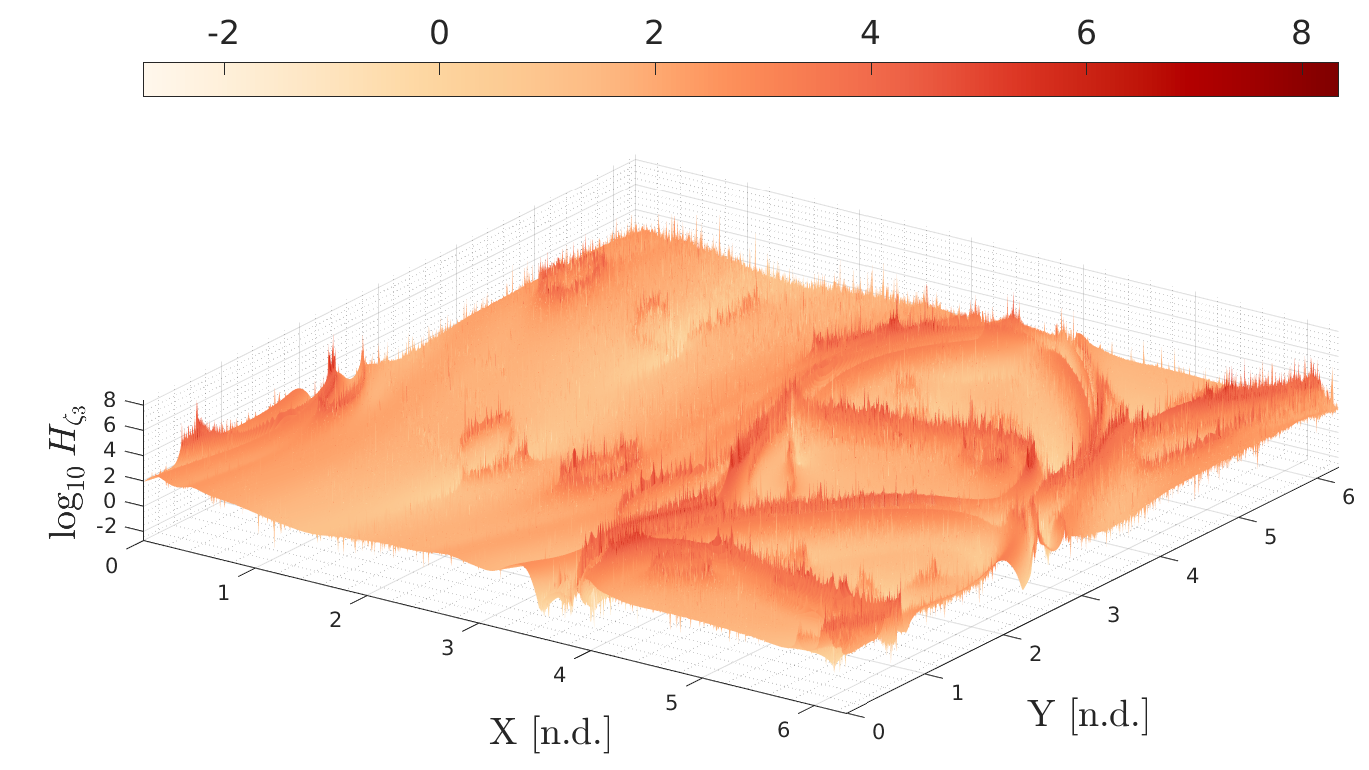}
\caption{Helicity field obtained using divided differences with auxiliary grid spacing of $0.1$ of the nominal grid spacing in all directions. The same grid is used for computing both $C^5_0$ and $\nabla\times\bm{\zeta}_3$.}
\label{f:forced_abc_helicity_dd}
\end{subfigure}
\caption{$-\log H_{\bm{\zeta}_3}$ for the chaotically-forced ABC flow from $t_0 = 0$ to $T = 5.0$. DA-LCS produces visibly better-defined ridges to identify seed points.}
\label{f:chaotic_abc_helicity}
\end{figure*}

Following \cite{Blazevski2014HyperbolicFlows}, we now demonstrate that DA-LCS is robust under perturbations from a chaotic forcing function $g\left(t\right)$. The motion is forced by a chaotic Duffing oscillator, with equations of motion given by
\begin{eqnarray}\label{eq:forcedabceomstart}
    \dot{x} &=& \left(A + 0.1\sin{t}\right)\sin{z} + C\cos{y} \\
    \dot{y} &=& B\sin{x} + \left(A + 0.1g\left(t\right)\right)\cos{z} \\
    \dot{z} &=& C\sin{y} + B\cos{x}\label{eq:forcedabceomend}
\end{eqnarray}
where $g\left(t\right)$ is the $x-$coordinate of the solution to the Duffing equation 
\begin{equation}\label{eq:eomduffingoscillator}
     \ddot{x} = -\delta \dot{x} - \beta x - \alpha x^3 + \gamma\cos\left(\omega t\right).
\end{equation}
with parameters $\alpha = 1$, $\beta = -1$, $\gamma = 0.3$, $\delta = 0.2$, $\omega = 1$.

The computational grid is again the same as for the previous test cases involving the ABC flow, including the hyperplanes $\mathcal{S} = \lbrace \left(x\,, y\,, z\right) \in \left[0,\,2\pi\right]^3~:~z=s_1\rbrace, s_1 = 0.0, 0.005, 0.01, \dots, 0.1$, but a longer integration time of $T = 5$ is used to match the literature. Again, a helicity tolerance of $\alpha = 5\times10^{-5}$ is used with a filtering distance of $d_F = 0.05$.

\begin{figure}
    \centering
    \includegraphics[height=0.3\textheight]{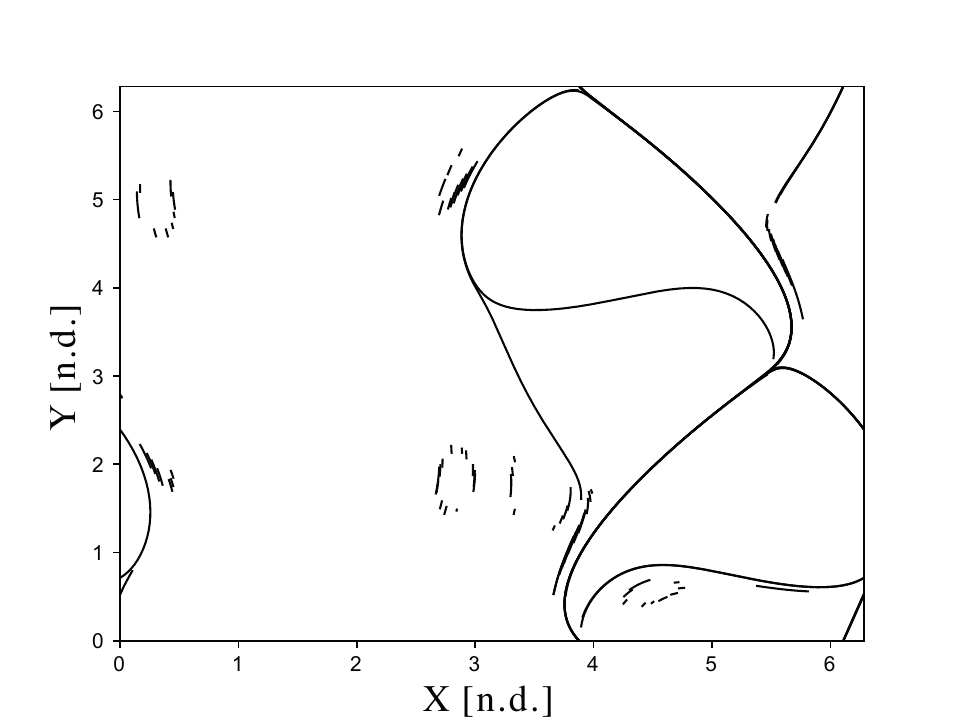}
    \caption{Final strainline structure on the $z = 0$ plane for the chaotically-forced ABC flow computed using DA-LCS. The structure is formed of $160$ individual strainline segments.}
    \label{f:chaotic_abc_final_lines}
\end{figure}

The FTLE fields computed using DA-LCS and divided differences are again shown in Figure \ref{f:forced_abc_ftle_da} and Figure \ref{f:forced_abc_ftle_dd}, respectively. The helicity fields are shown in Figures \ref{f:forced_abc_helicity_da} and \ref{f:forced_abc_helicity_dd}, respectively. The helicity field in particular now exhibits a significant difference compared to the two previous cases. Using DA-LCS, we are able to resolve a relatively smooth ridge of low helicity, whereas the use of divided differences leads to noticeable numerical noise throughout the field as well as an overall much higher helicity.

The strainlines for this system on the $z = 0$ plane computed using DA-LCS are presented in Figure \ref{f:chaotic_abc_final_lines}. A total of $160$ strainline segments give the full structure on the $z = 0$ plane.

\section{The Elliptic-Restricted Three-Body Problem}\label{sec:er3bp}

We now demonstrate the numerical out-performance of DA-LCS compared to standard approaches on a test problem from astrodynamics. The system presented in this Section is the Elliptic-Restricted Three-body Problem (ER3BP), which studies the motion of a small mass $m_3$ under the motion of two far larger masses $m_1$ and $m_2$ such that $m_1 \geqslant m_2 \gg m_3$. The system is parameterised by the mass parameter $\mu = m_2 / (m_1 + m_2)$.

In an inertial coordinate system, $m_2$ and $m_1$ orbit their centre of mass on an ellipse of fixed eccentricity $e_p$, which is the second system parameter. The angle of $m_2$ with respect to the $+x$-axis of the inertial coordinate system is the true anomaly $\nu$. 

For the special case of $e_p = 0$, one recovers an autonomous dynamical system for which fixed points and invariant manifolds exist \cite{Koon2008DynamicalDesign}; for the more general $e_p > 0$, such structures become difficult to determine. LCS have thus been suggested to analyse the behaviour for the cases of $e_p > 0$. 
In this example, we analyse the interesting dynamical phenomena around $m_2$. For small differences in initial position and velocity, orbits can vary from being bound entirely around $m_2$, being only temporarily captured around $m_2$, or escaping entirely \cite{Luo2015}. Profiling these regions is of high importance in the design of space missions \cite{Belbruno1987LunarMission}.

\begin{figure}
    \centering
    \def\svgwidth{0.75\textwidth}
    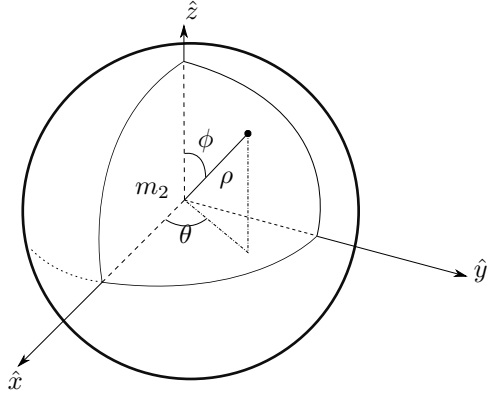
    \caption{The parameterisation of the space around $m_2$ using spherical coordinates relative to the inertial coordinate frame. By careful choice of the ranges of $\rho$, $\theta$ and $\phi$, the reference hyperplanes can encapsulate regions of `interesting' dynamics about $m_2$.}
    \label{f:spherical_parameterisation}
\end{figure}

Since the ER3BP lives in a phase space defined in $\mathbb{R}^6$, but the algorithm above functions for a CGST that is $3\times 3$ in dimension and represents a system with three-dimensional dynamics, we embed a three-dimensional submanifold in the six-dimensional phase space on which we compute the LCS. We parameterise the manifold in the three spatial directions to represent position around $m_2$ using spherical coordinates $\bm{\Psi} = \left(\rho,\theta,\phi\right)$ (Figure \ref{f:spherical_parameterisation}).
We complete the embedding by uniquely associating a velocity $\bm{v}$ with each point in space to complete the full phase space.

Given the Cartesian position $\bm{x}=\left(x,\,y,\,z\right)^\top$ corresponding to $\bm{\Psi}$
\begin{eqnarray}
	x &=& \rho\cos\theta\sin\phi\\
	y &=& \rho\sin\theta\sin\phi\\
	z &=& \rho\cos\phi
\end{eqnarray}
the velocity at this point $\bm{v}\left(\bm{x}\right)$ is chosen to be
\begin{equation}
	\bm{v}\left(\bm{x}\right) = \sqrt{\text{G}m_2\frac{\left(1+e\right)}{\rho^3}}\left[
	\begin{pmatrix}x\\y\\z\end{pmatrix}\times\begin{pmatrix}0\\0\\1\end{pmatrix}\right],
\end{equation}
where the problem parameters $\text{G}m_2$ and $e$ are the gravitational parameter of $m_2$ and an orbital eccentricity, respectively. Conceptually, this fixes the velocity direction tangential to a cylinder around the $z$-axis, while the magnitude corresponds to a Keplerian orbit of eccentricity $e$ around $m_2$. Together, this choice of velocity vector reveals the `dynamically interesting' behaviour introduced previously.

Rather than using the inertial coordinate system about $m_2$ to propagate the initial condition, it is beneficial to use a rotating-pulsating Cartesian coordinate system centred on the barycentre of $m_1$ and $m_2$. In this system, $m_1$ and $m_2$ are fixed, and the true anomaly $\nu$ replaces time as the independent variable. The transformation of the initial condition into this coordinate system is shown in \ref{sec:app:er3bp_derivation}. In this system the equations of motion are given by
\begin{eqnarray}
x^{\prime\prime} &=& 2y^\prime + \pdiff{\Omega}{x}\\
y^{\prime\prime} &=& -2x^\prime + \pdiff{\Omega}{y}\\
z^{\prime\prime} &=& \pdiff{\Omega}{z}
\end{eqnarray}
where
\begin{equation}
    \Omega = \frac{1}{1 + e_p\cos\nu}\left[ \frac{1}{2}\left(x^2 + y^2 - z^2 e\cos\nu\right) + \frac{\mu}{r_1} + \frac{1-\mu}{r_2} + \frac{1}{2}\mu\left(1-\mu\right)\right]
\end{equation}
and
\begin{eqnarray}
    r_1 &=& \sqrt{\left(x - \mu\right)^2 + y^2 + z^2}\\
    r_2 &=& \sqrt{\left(x + 1 - \mu\right)^2 + y^2 + z^2}.
\end{eqnarray}

After propagation under the equations of motion, the transformation into the rotating coordinate system is inverted, and the final position is projected back into spherical coordinates. Another advantage of DA-LCS is that, provided the intermediate transformations are coded as DA operations, the derivatives of this process are computed fully automatically and there is no need to derive further equations for the coordinate transformations. 

For this example, we choose $m_1$ to be the Sun and $m_2$ to be Mars, with the system parameters as given in Table \ref{tab:er3bp_parameter_values}. The set of reference hyperplanes is defined as $$\mathcal{S} = \left\{\bm{\Psi}\in\left[r, r_s\right]\times\left[0,2\pi\right]\times
\left[5^\circ,\,15^\circ,\dots,175^\circ\right]\right\}.$$
The variables $r$ and $r_s$ here are the radius and the Hill sphere of Mars, respectively; the latter is the maximum distance from Mars at which it still dominates gravitational attraction. Together, the reference planes enclose the `dynamically interesting' region around $m_2$. The initial integration time is set equal to $t_0 = \nu_0 = 0$ and the final time is $T = \nu = 2\pi$. The helicity tolerance $\alpha$ used is $10^{-5}$.

\begin{figure*}
    \centering
    \begin{subfigure}{0.75\textwidth}
        \includegraphics[width=\linewidth]{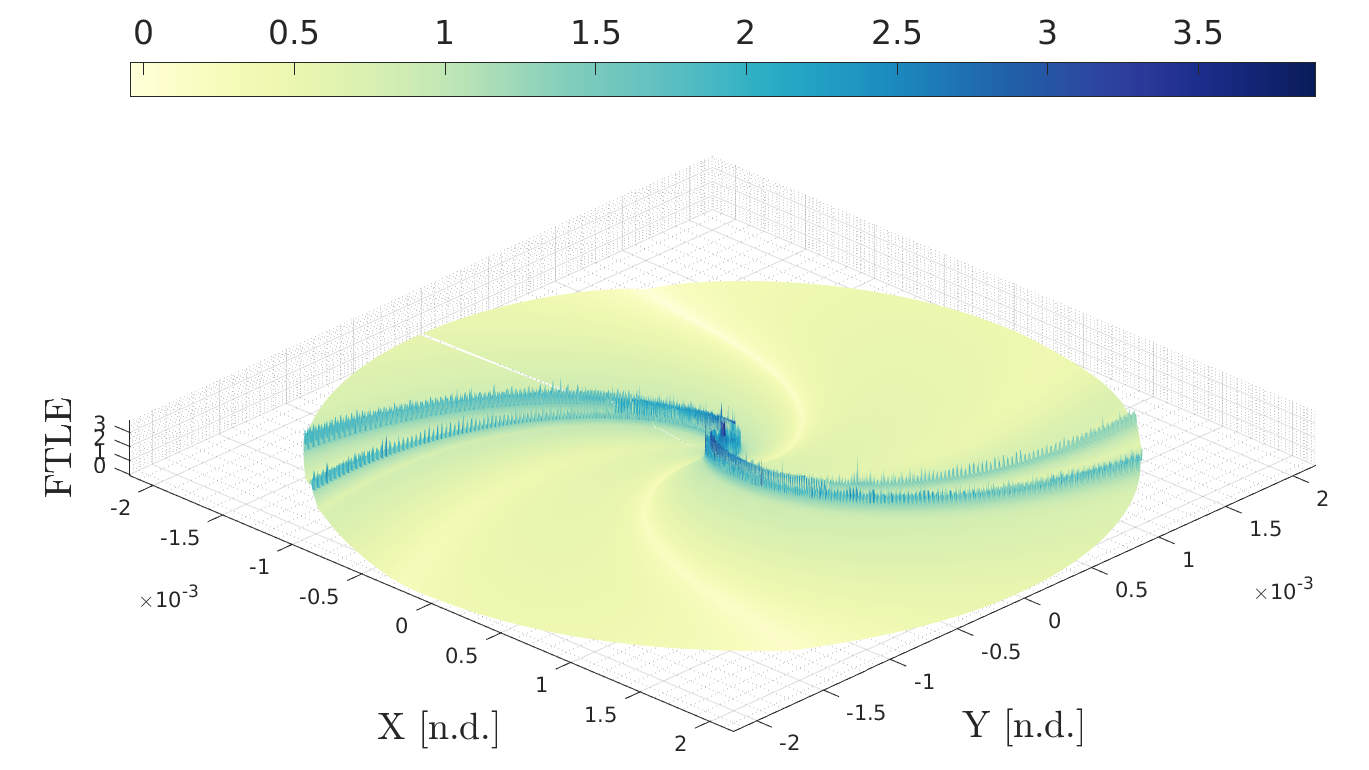}
        \caption{Computed using DA-LCS.}
        \label{f:er3bp_ftle_da}
    \end{subfigure}\hfill
    \begin{subfigure}{0.75\textwidth}
        \includegraphics[width=\linewidth]{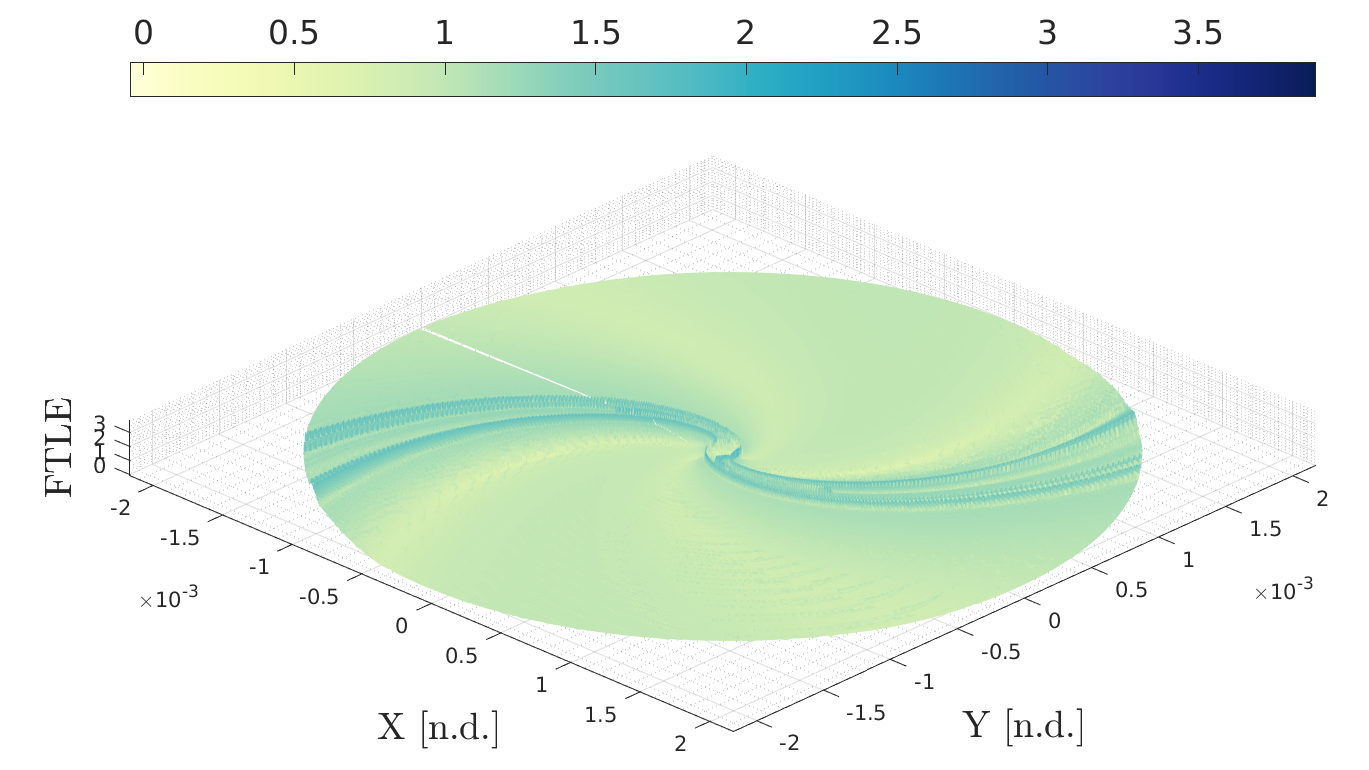}
        \caption{FTLE field obtained using divided differences with auxiliary grid spacing of $0.05$ of the nominal grid spacing in $r$ and $\phi$ and the nominal grid spacing in $\theta$.}
        \label{f:er3bp_ftle_dd}
    \end{subfigure}
    \caption{Finite-time Lyapunov exponent field for the Elliptic-Restricted Three-body Problem on the $\theta = 115^\circ$ plane from $t_0 = \nu_0 = 0$ to $T = \nu = 2\pi$. While the structure is qualitatively the same, the ridges in the FTLE field are much more well-defined with DA-LCS.}
\end{figure*}
\begin{figure*}
    \centering
    \begin{subfigure}{0.75\textwidth}
        \includegraphics[width=\linewidth]{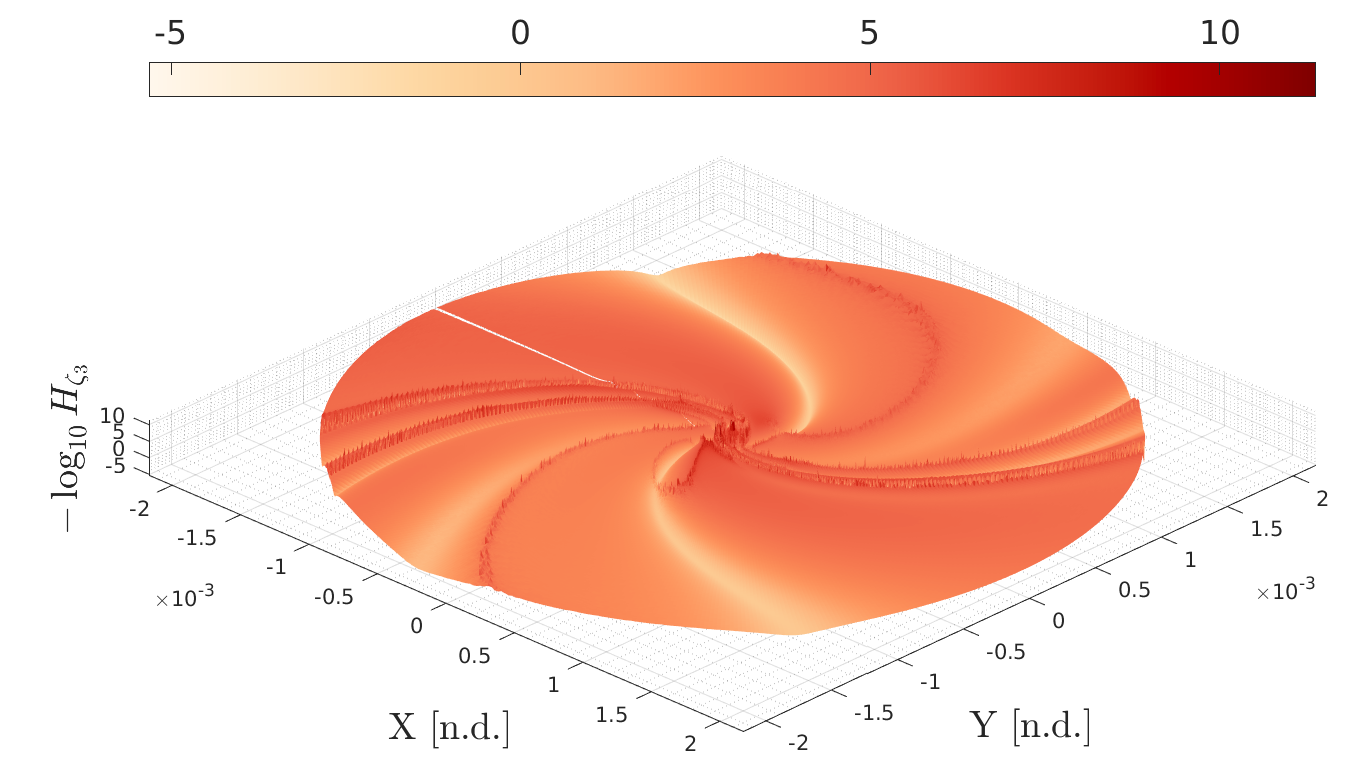}
        \caption{Computed using DA-LCS.}
        \label{f:er3bp_helicity_da}
    \end{subfigure}\hfill
    \begin{subfigure}{0.75\textwidth}
        \includegraphics[width=\linewidth]{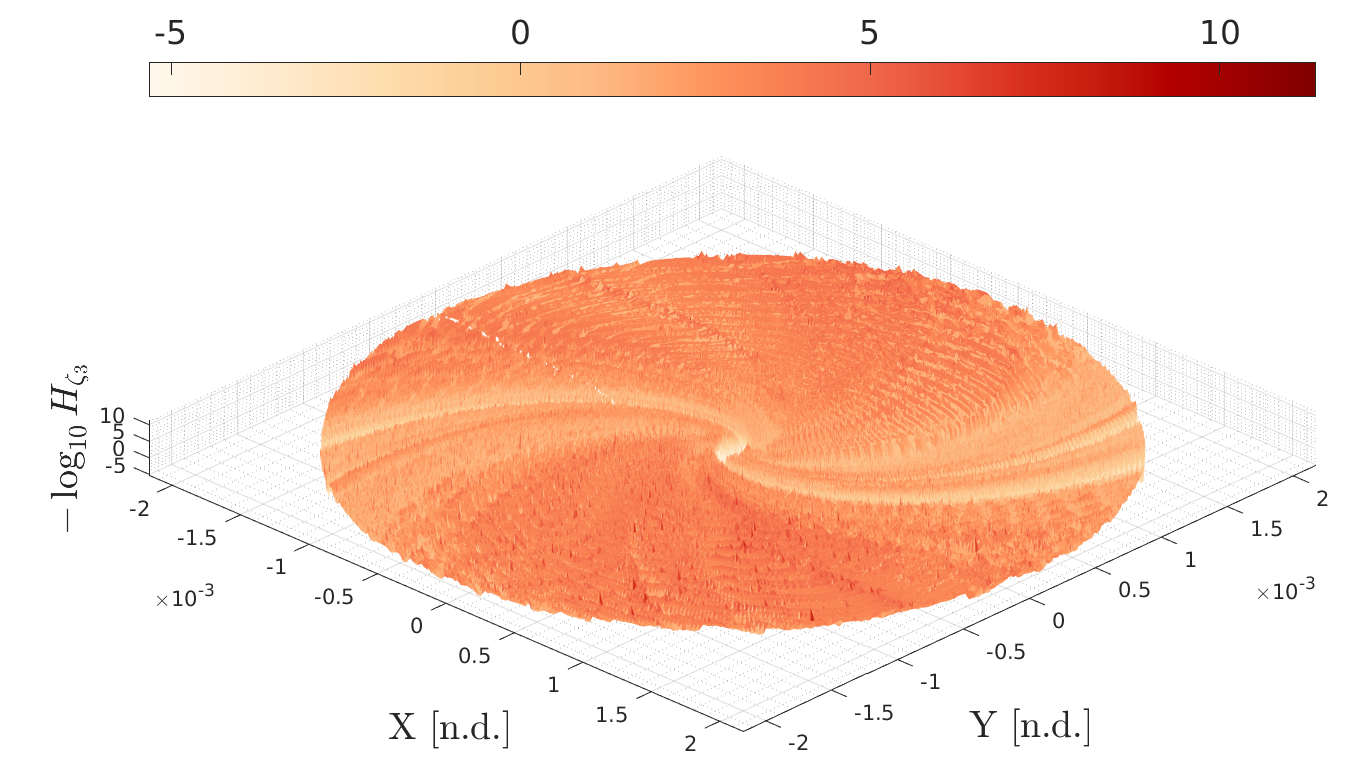}
        \caption{Helicity field obtained using divided differences with auxiliary grid spacing of $0.05$ of the nominal grid spacing in $r$ and $\phi$ and the nominal grid spacing in $\theta$. The same grid is used for computing both $C^{2\pi}_0$ and $\nabla\times\bm{\zeta_3}$.}
        \label{f:er3bp_helicity_dd}
    \end{subfigure}
    \caption{$-\log{H_{\bm{\zeta}_3}}$ for the Elliptic-Restricted Three-body Problem on the $\theta = 115^\circ$ plane from $t_0 = \nu_0 = 0$ to $T = \nu = 2\pi$. No defined regions of low helicity are visible when using divided differences, whereas with DA-LCS we can readily identify low helicity regions to identify seed points.}
\end{figure*}

\begin{table*}[]
    \centering
    \begin{tabular}{lcr}
    \textbf{Parameter} & \textbf{Description} & \textbf{Value}\\\hline
    $e_p$ & Eccentricity of the orbit of $m_2$ about $m_1$ & $0.0935$\\
    $\mu$ & Mass parameter & $3.227154\times10^{-7}$\\
    $e$ & Eccentricity of the orbit of $m_3$ about $m_2$ & $0.9$\\
    $\text{G}m_2$ & Standard gravitational parameter of $m_2$ & $1.50499\times 10^{-14}$\\
    $r$ & Planetary radius of $m_2$ & $1.641\times 10^{-5}$ \\
    $r_s$ & Hill sphere of $m_2$ & $0.00513$\\
    \end{tabular}
    \caption{Parameter values used in the ER3BP investigation where $m_1$ is arbitrarily chosen to be the Sun and $m_2$ arbitrarily chosen to be Mars. All values are given in non-dimensional units and valid at $\nu = 2n\pi,\,n\in\mathbb{Z}$.}
    \label{tab:er3bp_parameter_values}
\end{table*}

\subsubsection{Results}

\begin{figure}
    \centering
    \includegraphics[height=0.3\textheight]{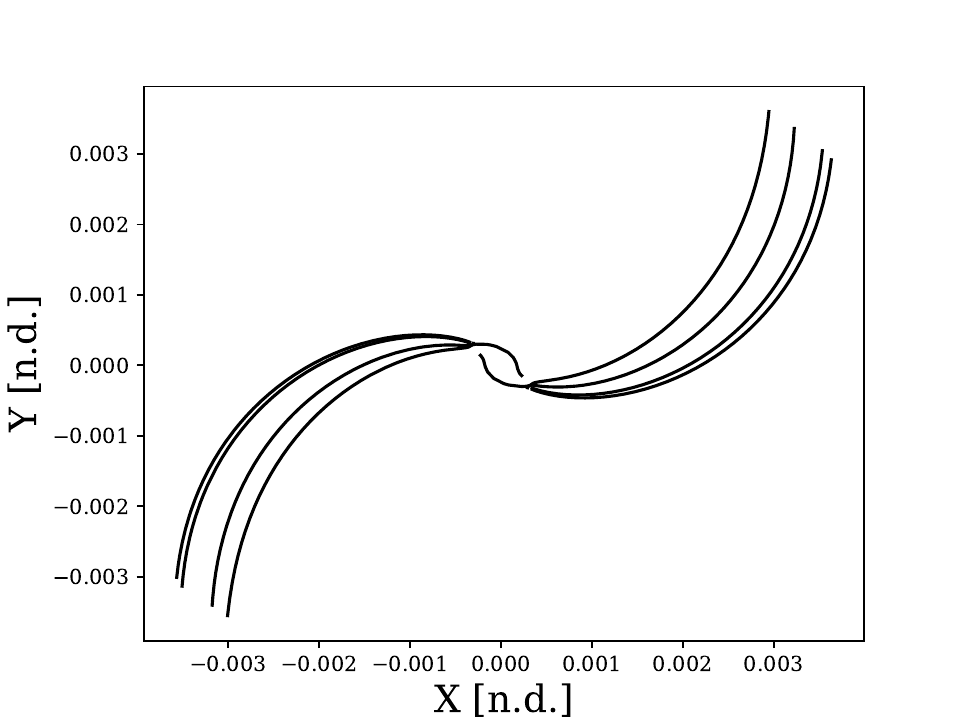}    
    \caption{Strainlines on the $\theta = 115^\circ$ plane for the Elliptic-Restricted Three-Body Problem computed using DA-LCS. We are unable to generate any strainlines when using divided differences, but with DA-LCS we can deduce the structure of the LCS readily and with only 8 strainlines.}
    \label{f:strainlines_er3bp_da}
\end{figure}

The FTLE fields computed using DA-LCS and divided differences on the $\theta = 115^\circ$ plane is presented in Figures \ref{f:er3bp_ftle_da} and \ref{f:er3bp_ftle_dd}, respectively. The structure found using DA-LCS agrees with what would be expected from previous literature, with the structures in the two `arms' being consistent with the transition between orbits that escape and are permanently or temporarily captured about $m_2$ \cite{Luo2015}. Similar performance, albeit with poorer definition of the FTLE ridges, can be obtained using divided differences after tuning the grid sizes used to generate the derivatives. We note that the ER3BP does admit variational equations that can be integrated with the equations of motion which may improve the quality of the derivatives used to compute $\cgst$.

Importantly, these variational equations cannot be used to compute $\nabla\times\bm{\zeta}_3$, which must still be approximated using divided differences and appear to produce the majority of the error for this test case. This is to be expected, as the estimation of second derivatives using divided differences is numerically difficult. Figure \ref{f:er3bp_helicity_da} presents the helicity field on the $\theta = 115^\circ$ plane for the ER3BP computed using DA-LCS, which like the FTLE field highlights the `arms' as being influential portions of flow. Qualitative inspection of the trajectories in this region reveals the low-helicity portions of the field to separate regions of different dynamical behaviour. However, using divided differences to compute the helicity, given in Figure \ref{f:er3bp_helicity_dd}, produces no meaningful insight into the helicity field even after tuning the grid sizes used; \revadd{the numerical noise in the determination of the helicity reveals no distinct ridges along which the numerical integration can begin, and the accuracy of the eigenvectors of $\cgst$ when using divided differences yields strainlines that do not follow the expected structure in previous attempts at this topic \cite{RosRoca2015ComputationBoundaries, Parkash2019ApplicationTrajectories}, even after extensive tuning of the grid size used.} This numerical improvement comes completely automatically, without the need to tune grid sizes and functions without any \textit{a priori} knowledge.

\begin{figure}
    \centering
    \begin{subfigure}{.6\textwidth}
        \includegraphics[width=\linewidth]{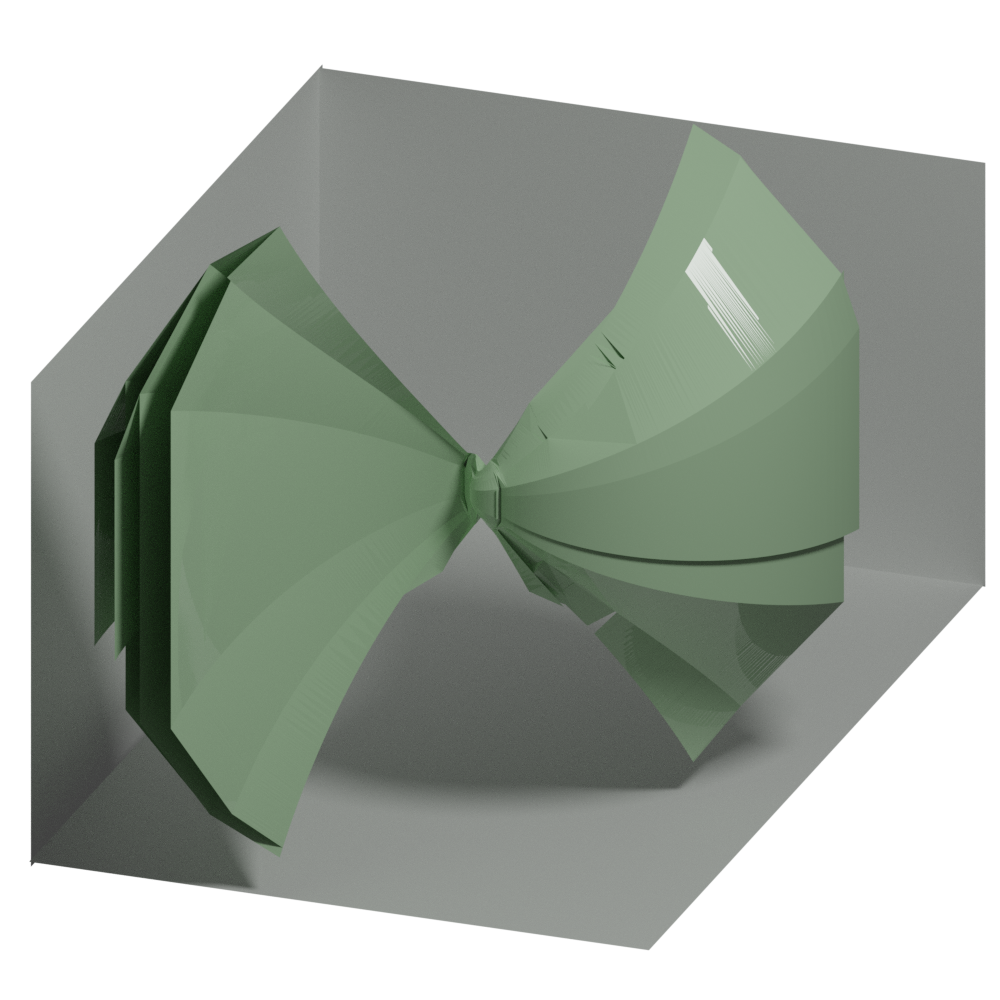}
        \caption{Full 3D structure of the LCS over the entire set of reference planes.}
    \end{subfigure}\hfill%
    \begin{subfigure}{.6\textwidth}
        \includegraphics[width=\linewidth]{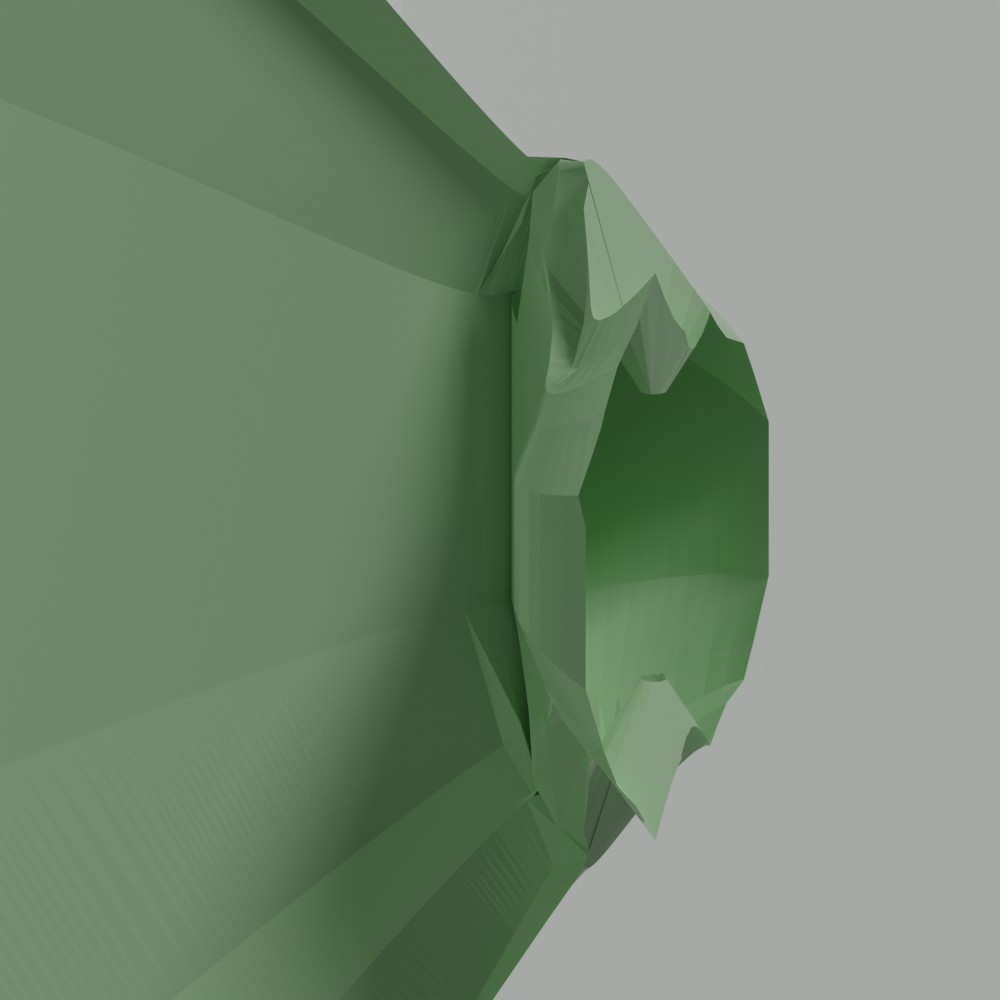}
        \caption{A zoomed-in section of the full LCS highlighting the interior structure.}
    \end{subfigure}%

    \caption{A set of representative renders of the 3D LCS for the ER3BP test case. The left figure is the full 3D LCS over all hyperplanes in $\mathcal{S}$. On the right is a zoomed-in portion of the centre of the LCS, with the right half removed to highlight the internal structure.}
    \label{f:er3bp_rendering}
\end{figure}

The final strainlines for this flow computed using DA-LCS on the $\theta = 115^\circ$ plane are shown in Figure \ref{f:strainlines_er3bp_da}, and largely follow from the helicity field given earlier. We were not able to generate any meaningful strainlines using divided differences due to the poor numerical resolution of the eigenvectors and the related helicity field. A representative rendering of the full 3D LCS for this test case is shown in Figure \ref{f:er3bp_rendering}.

\section{Conclusion}

This paper has introduced DA-LCS, an improved numerical method for determining hyperbolic Lagrangian Coherent Structures in time-dependent dynamical systems. We showed how Differential Algebra can be used to directly construct high-order Taylor expansions of the flow, its derivatives and a field of leading eigenvectors of the flow's strain tensor, accurate to machine precision. We have shown that with this information we can construct a highly-accurate LCS based solely on the underlying dynamics of the system, even in highly complex flows. We demonstrated the effectiveness of the method through applications to common variations of the Arnold-Beltrami-Childress flow from the literature, as well as introducing a new and particularly challenging test problem from astrodynamics where the classical methods fail to produce usable results. DA-LCS also constructs the LCS automatically and without any \textit{a priori} information, requiring no additional implementation beyond the dynamics of the system.

\section*{Acknowledgements}

The authors acknowledge financial support from the EPSRC Centre for Doctoral Training in Next Generation Computational Modelling grant EP/L015382/1, and the use of the IRIDIS High Performance Computing Facility and associated support services at the University of Southampton. The authors also thank Davide Lasagna for his helpful suggestions.


 

\newpage

\bibliography{references}

\appendix
\cleardoublepage

\begin{figure}
    \centering
    \def\svgwidth{\linewidth}
    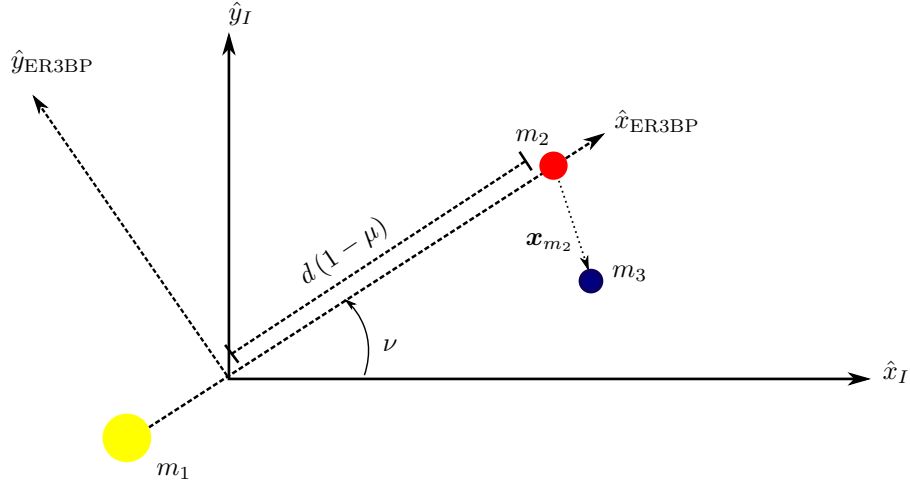
    \caption{Schematic of the inertial frame (subscript $_I$) and the rotating-pulsating frame (subscript $_\text{ER3BP}$) for use in Appendix \ref{sec:app:er3bp_derivation}. The transformation between the inertial and rotating-pulsating frame is a composite translation, rotation and normalisation.}
    \label{f:problem_setup}
\end{figure}

\section{Transformation into the rotating-pulsating frame of the Elliptic-Restricted Three-body Problem}\label{sec:app:er3bp_derivation}

As previously introduced, the Elliptic-Restricted Three-body Problem (ER3BP) models the motion of a small object $m_3$ under the influence of two far larger masses $m_1$ and $m_2$, such that $m_1 \geqslant m_2 \gg m_3$. The object $m_3$ is sufficiently small compared to $m_1$ and $m_2$ that it is considered massless. The system is parameterised by the mass parameter $\mu = m_2 / \left(m_1 + m_2\right)$, and in an inertial coordinate system $m_1$ and $m_2$ orbit their center of mass on an ellipse with fixed eccentricity $e_p$. 

In Section \ref{sec:er3bp}, we chose the parameterisation of the sub-manifold to represent initial position around $m_2$ in the inertial frame using spherical coordinates, and the embedding to represent the initial velocity of the point in the inertial frame. This was done to simplify the problem set-up and more easily define the regions of `interesting' dynamical behaviour. However, in the literature \citep{Szebehely1968} the ER3BP is integrated in a rotating coordinate system where $m_1$ and $m_2$ are fixed on the $x-$axis at $\left(-\mu,0,0\right)$ and $\left(1-\mu,0,0\right)$, respectively, and the distance between them is normalised to unity. In this frame, the independent variable in the motion of $m_3$ is the true anomaly $\nu$. To simplify the test case, the transformation that follows is valid only for values of $\nu$ that are scalar multiples of $2\pi$; for an in-depth derivation of the general case of this transformation, the reader is directed to \cite{Szebehely1968}.

With reference to Figure \ref{f:problem_setup}, the transformation of the position from the $m_2$-centred inertial frame to the rotating-pulsating frame is formed of a translation to move the centre of the system to the centre of mass of $m_1$ and $m_2$, a rotation to align the $+x$ axis to the line joining $m_1$ and $m_2$, and a scaling to normalise the distance between $m_1$ and $m_2$ to unity.

We perform the translation first. Define the Cartesian position of $m_3$ about $m_2$ in the inertial frame as $\bm{x}_{m_2}$, such that the translated position around the barycentre (centre of mass) of $m_1$ and $m_2$, $\bm{x}_\text{BC}$, is
\begin{equation}
	\bm{x}_\text{BC} = \bm{x}_{m_2} + d\left(1-\mu\right)\begin{pmatrix}
	\cos\nu \\
	\sin\nu \\
	0
	\end{pmatrix}
\end{equation}
where $d$ is the full distance between $m_1$ and $m_2$, and $\left(1-\mu\right)$ gives the proportion of the distance $d$ between $m_2$ and the centre of mass. The distance $d$ can be retrieved from the orbit equation (more generally known as the ellipse equation)
\begin{equation}
d\left(\nu\right) = \frac{a\left(1 - e_p^2\right)}{1 + e_p\cos\nu}
\end{equation}
with $a$ the semi-major axis of $m_2$ about $m_1$. For the case of $m_1$ being the Sun and $m_2$ being Mars studied in this paper, at scalar multiples of $2\pi$ the semi-major axis $a = 1.10314$.

The coordinate axes must now be rotated such that $m_1$ and $m_2$ lie on the $+x$-axis. This is a clockwise rotation about $+z$ of an angle $\nu$. We apply the standard Euler rotation matrix to $\bm{x}_\text{BC}$ to find its equivalent state in the rotated coordinate system $\bm{x}_\text{rot}$
\begin{equation}
\bm{x}_\text{rot} = R_z\left(\nu\right)\bm{x}_\text{BC} = \begin{pmatrix}
\cos\nu & \sin\nu & 0\\
-\sin\nu & \cos\nu & 0\\
0 & 0 & 1
\end{pmatrix}
\bm{x}_\text{BC}.
\end{equation} 

Finally, the distance between $m_1$ and $m_2$ is normalised to $1$ by scaling the length unit of the system by $d$. This yields the final ER3BP position $\bm{x}_\text{ER3BP}$
\begin{equation}
\bm{x}_\text{ER3BP} = \frac{\bm{x}_\text{rot}}{d}.
\end{equation}
The composite transformation can be combined into a single equation for brevity:
\begin{eqnarray}
\bm{x}_\text{ER3BP} &=& \frac{R_z\left(\nu\right)}{d\left(\nu\right)}\left(\bm{x}_{m_2} + d\left(1-\mu\right)
\begin{pmatrix}
\cos\nu\\
\sin\nu\\
0
\end{pmatrix}
\right) \\
&=& \frac{R_z\left(\nu\right)}{d\left(\nu\right)}\bm{x}_{m_2} + \left(1-\mu\right)
\begin{pmatrix}
1\\
0\\
0
\end{pmatrix}\label{eq:app:positiontransformation}.
\end{eqnarray}
The equation above completes the transformation of the position from the inertial coordinate system around $m_2$ to the rotating coordinate system of the ER3BP. However, integrating the ER3BP equations of motion also requires the initial velocity of $m_3$ with respect to $\nu$ in the rotating coordinate system. Thus, the velocity in the inertial frame about $m_2$  with respect to time given by the embedding introduced in the main text, $\bm{v}$, must also be transformed into the ER3BP coordinate frame.

To do this, Equation \ref{eq:app:positiontransformation} is differentiated with respect to the true anomaly $\nu$, which is the independent variable in the ER3BP. In the following, $\Box^\prime$ denotes derivatives with respect to $\nu$ (as in the ER3BP coordinate system), and $\dot{\Box}$ denotes derivatives with respect to time (the inertial coordinate system.) Via the chain rule, the derivative of Equation \ref{eq:app:positiontransformation} is
\begin{equation}
    \bm{x}^\prime_\text{ER3BP} = \frac{R_z\left(\nu\right)^\prime}{d\left(\nu\right)}\bm{x}_{m_2} + \frac{R_z\left(\nu\right)}{d\left(\nu\right)}\bm{x}^\prime_{m_2}
\end{equation}
since the quantity $\left(1/d\left(\nu\right)\right)^\prime$ is zero in the case of $\nu$ being a scalar multiple of $2\pi$.
The quantity $R_z\left(\nu\right)^\prime$ is trivial to infer from its use previously
\begin{equation}
    R_z^\prime\left(\nu\right) = \begin{pmatrix}
    -\sin\nu & \cos\nu & 0 \\
    -\cos\nu & -\sin\nu & 0 \\
    0 & 0 & 0
    \end{pmatrix}.
\end{equation}
The velocity with respect to time in the inertial frame $\bm{v}$ represents $\dot{\bm{x}}_{m_2}$. To obtain $\bm{x}^\prime_{m_2}$, we use
\begin{equation}
    \diff{\bm{x}_{m_2}}{\nu} = \diff{\bm{x}_{m_2}}{t}\diff{t}{\nu} = \bm{v} / \dot{\nu}
\end{equation}
where $\dot{\nu}$ is given by considering the angular momentum of $m_2$ about $m_1$
\begin{equation}
    \dot{\nu} = \frac{\text{G}m_1^\frac{1}{2}\left(1 + e_p\right)^2}{a^\frac{3}{2}\left(1-e_p^2\right)^\frac{3}{2}}
\end{equation}
which completes the transformation of a position in the inertial frame about $m_2$ to the rotating coordinate system of the ER3BP for use in Section \ref{sec:er3bp}.

Since we are computing the LCS on a submanifold that represents the spatial dimensions about $m_2$, the inverse transformation need only consider the position. Equation \ref{eq:app:positiontransformation} is inverted to give $\bm{x}_{m_2}$ and then converted back into spherical coordinates for use in computing the LCS.

\linenumbers

\end{document}